\documentclass[fleqn,usenatbib]{mnras}

\usepackage[T1]{fontenc}
\usepackage{ae,aecompl}

\usepackage{graphicx}	
\usepackage{amsmath}	
\usepackage{amssymb}	
\usepackage{threeparttable}
\usepackage{newtxtext,newtxmath}

\newcommand{\msun}{M_{\odot}}

\title[Massive accretions influence infall of satellites]{The infall of dwarf satellite galaxies are influenced by their host's massive accretions}

\author[D'Souza \& Bell]{Richard D'Souza$^{1}$ $^{2}$, 
\thanks{Contact e-mail:\href{mailto:rdsouza@speola.va}{rdsouza@specola.va}}
						Eric F.\ Bell$^{2}$\\
$^{1}$Vatican Observatory, Specola Vaticana, V-00120, Vatican City State\\
$^{2}$University of Michigan, Department of Astronomy, 311 West Hall, 1085 South University Ave., Ann Arbor, MI 48109-1107, USA.
}

\date{Accepted 2021 April 27. Received 2021 April 6; in original form 2021 January 4}

\pubyear{2021}

\begin{document}
\label{firstpage}
\pagerange{\pageref{firstpage}--\pageref{lastpage}}
\maketitle

\begin{abstract}
Recent progress in constraining the massive accretions (>1:10) experienced by the Milky Way (MW) and the Andromeda galaxy (M31) offers an opportunity to understand the dwarf galaxy population of the Local Group. Using zoom-in dark matter-only simulations of MW-mass haloes and concentrating on subhaloes that are thought to be capable of hosting dwarf galaxies, we demonstrate that the infall of a massive progenitor is accompanied with the accretion and destruction of a large number of subhaloes. Massive accreted progenitors do not increase the total number of infalling subhaloes onto a MW-mass host, but instead focus surrounding subhaloes onto the host causing a clustering in the infall time of subhaloes. This leads to a temporary elevation in the number of subhaloes as well as changes in their cumulative radial profile within the virial radius of the host. Surviving associated subhaloes with a massive progenitor have a large diversity in their orbits. We find that the star formation quenching times of Local Group dwarf spheroidal galaxies ($10^{5} \mathrm{\msun} \lesssim \mathrm{M}_{*} \lesssim 10^{7} \mathrm{\msun}$) are clustered around the times of the most massive accretions suffered by the MW and M31. Our results imply that a) the quenching time of dwarf spheroidals is a good proxy of their infall time and b) the absence of recently quenched satellites around M31 suggests that M33 is not on its first infall and was accreted much earlier.
\end{abstract}




\section{Introduction}
Improved observational data of dwarf satellite galaxies, especially in the Local Group, are key to resolving many of the pressing small-scale challenges associated with the nature of dark matter \citep[e.g.][]{Bullock2017}. Although the Milky Way (MW) and the Andromeda galaxy (M31) are similar in size, their satellite populations have a number of striking and unexplained differences. Out to a projected distance of 150 kpc, M31 has double the number of satellites as that of the MW.  Apart from the difference in numbers, there are also differences in the distributions of the satellite populations. The radial distribution of the satellites of the MW is much more centrally concentrated than that of the satellites of M31, and is possibly at tension with predictions from simulations \citep[e.g.][]{Samuel2020a, Carlsten2020}. A large fraction of M31's satellites ($\sim$ 13 out of the 27) are co-rotating in a thin plane in the sky \citep[RMS $\sim$ 14.1 kpc;][]{Ibata2014} hinting to common dynamical orbital properties and direction of angular momentum, and are found preferentially on the near side of M31, closer to the MW. In the MW, the distribution of orbital poles of 7 of the 11 classical satellites is strongly clustered in a direction normal to the disk of the galaxy, while Sculptor orbits along the same plane but in the opposite direction \citep[e.g.][]{Pawlowski2020}. Equally intriguing are the number of large streams discovered in the halo of M31, possibly caused by the recent destruction of satellites comparable or larger than the progenitor of the Sagittarius stream, prompting suggestions that M31 suffered up to $\sim$ 5 different accretion events in the last 3 or 4 Gyr \citep{McConnachie2018}. Preliminary measurements of the star formation histories of M31's satellites from resolved HST data reaching down to the red clump stars suggest that there are large differences in the quenching lookback times of the satellite populations of the MW and M31 \citep{Skillman2017,Martin2016,Weisz2019b}. While upcoming data through new projects like the Rubin Observatory \citep{Ivezic2019}, Subaru's Hyper Suprime-Cam and its prime focus spectrograph as well the Nancy Grace Roman Space Telescope should better inform our understanding of these problems, the differences in the satellite populations of the MW and M31 and their deviations from the mean expectations of the LCDM model raise a number of interesting questions. Chief among them is whether the differences in their satellite populations can be attributed to the differences in the accretion histories of the MW and M31.

The Large Magellanic Cloud (LMC), currently believed to be on its first infall into the MW \citep{Besla2010}, is expected to have contributed a number of dwarf galaxies to the Galaxy's existing satellite population \citep[e.g.][]{Deason2015}. A number of ultra-faint dwarf galaxies have been already spatially and kinematically associated with the LMC \citep[e.g.][]{Sales2011,Sales2017,Kallivayalil2018}, while there are hints that a few classical dwarfs may also have been accreted with a more massive LMC (e.g. Fornax and Carina, \citealt{Pardy2019,Erkal2019}, although see \citealt{Patel2020} for a different view). The infall of M33 onto M31 \citep{McConnachie2009} is also expected to have contributed a number of dwarf galaxies to the existing satellite population of M31 \citep[e.g.][]{Patel2018}. Yet, similar to these ongoing massive accretion events (>1:10), it is reasonable to expect that the past massive accretions of the MW and M31 would also have contributed significant number of dwarf satellites to their satellite populations.

Today, we have much better constraints on the past massive accretion events of both the MW and M31. Thanks to Gaia, it is now believed that the MW accreted a Small Magellanic Cloud-like galaxy $\sim$ 10 Gyr ago \citep[Gaia-Enceladus-Sausage;][]{Helmi2018,Belokurov2018}. On the other hand, M31's large metal-rich stellar halo \citep{Ibata2014} containing intermediate age stars \citep{Brown2006,Brown2007} indicates that it merged with a large metal-rich galaxy ($\log \mathrm{M}_{*} \sim 10.3$) nearly $\sim$2 Gyr ago \citep{DSouza2018b}. Such a recent merger scenario is also independently supported by simulations which try to reproduce the steep age-velocity dispersion \citep{Dorman2015} of M31's stellar disk \citep{Hammer2018}. Given the large differences in their massive accretion histories of the MW and M31, one might expect appreciable differences also in their satellite populations.

The infall time of the dwarf satellites (when they enter into the radius of their host's virial halo) is an essential ingredient in deciphering their association with past massive accretions of the MW and M31. However, constraining the infall time of dwarf satellites in the Local Group remains far from being an easy task. Even if one can constrain the full phase-space information of the dwarf satellites, retracing their orbital histories is plagued with a number of difficulties. Apart from the uncertainties in the present-day potential, it is doubly difficult to account for the time variation in the potential owing to the MW's and M31's massive (>1:10) accretion events. This is expected to prevent an easy inference of the infall time of the satellites from their respective binding energies \citep[e.g.][]{Rocha2012} and complicate a backward integration to retrace the infalling orbits of the satellites (e.g., \citealp{Patel2020}, although see \citealp{Vasiliev2021} for encouraging progress on this front; see also D'Souza et al. in preparation).

Improvements in the measurements of the shutdown time of star formation in classical dwarf spheroidal satellite galaxies of the Local Group \citep{Weisz2015,Weisz2019b} offers us a complementary way of constraining the infall time of dwarf spheroidal satellites.  While the precise mechanisms governing the shutdown of star formation in dwarf spheroidal satellite galaxies are not properly understood, it is generally believed that classical dwarf spheroidal galaxies \citep[$10^{5} \mathrm{\msun} \lesssim \mathrm{M}_{*} \lesssim 10^{7} \mathrm{\msun}$;][]{Bullock2017} are quenched in the environment of their MW-mass hosts. On the other hand, brighter dwarfs \citep[$10^{7} \mathrm{\msun} \lesssim \mathrm{M}_{*} \lesssim 10^{9} \mathrm{\msun}$;][]{Bullock2017} may continue forming stars even after infall into the halo of their host galaxy \citep{Geha2012,Slater2014,Geha2017}, while the early quenching time of ultra-faint dwarfs \citep[$\mathrm{M}_{*} \lesssim 10^{5} \mathrm{\msun}; M_{\mathrm{V}} > - 7.7$;][]{Bullock2017,Simon2019} is suggestive that this class of galaxies had their star formation shut off by reionization (by $z \sim 6$). The high quenching fraction of classical dwarf spheroidals with respect to brighter dwarfs suggests a much shorter quenching time ($\sim$ 1 Gyr) for classical dwarfs \citep{Slater2014,Fillingham2015}. At least for classical dwarf spheroidals, their quenching lookback times may help us constrain their infall times.

Within the context of the LCDM paradigm, because it has been historically impossible to resolve statistical samples of faint dwarf satellites using hydrodynamical models, the infall of satellites has been traditionally studied instead using subhaloes capable of hosting dwarf satellites. It has been shown that the abundance of subhaloes associated with a dark matter halo is a strong function of the mass of the halo \citep[e.g.][]{Kravtsov2004}. However at a fixed halo mass, there is considerable scatter in the number of associated subhaloes \citep[super-Poissonian;][]{Boylan2010}. Subhaloes which are accreted early are likely to merge with the main galaxy, while subhaloes accreted later continue to survive till the present day \citep[e.g.][]{Font2006,Sales2007}. This leads to a correlation between the number of subhaloes found within its virial radius and the formation history of the halo \citep{Zentner2005,Zhu2006,Ishiyama2008}. A significant fraction of the subhaloes that were accreted by the MW-mass host are also presently found beyond the virial radius \citep[which we will term splashback galaxies in this paper, following, e.g.][]{Knebe2011, Teyssier2012}\footnote{While the term `splashback galaxies' was not intended to include galaxies that escape the potential of the main galaxy entirely, our radius-based definition would include such escaping galaxies as `splashback galaxies'.}. No differences were found the abundance or kinematics of substructure within the virial radii of isolated versus paired MW-mass haloes \citep{Garrison2014}. The ongoing accretion of a massive progenitor (>1:10; like in the case of the LMC) is expected to contribute a significant number of subhaloes hosting dwarf satellites to a MW-mass host \citep[e.g.][]{Sales2013,Deason2015,Dooley2017,Pardy2019}. However, we lack a systematic study of how the contribution of subhaloes through massive accretions (past as well as ongoing) can affect the present-day satellite populations of MW-mass galaxies.

In this paper, we focus on building physical and statistical intuition about how the infall properties of subhaloes hosting dwarf satellites ($M_{\mathrm{V}} > -8$) are correlated with the accretion of massive progenitors in a MW-mass halo. For this, we use a suite of 48 high resolution dark-matter only simulations of individual MW-mass haloes from the Exploring the Local Volume in Simulations \citep[ELVIS;][]{Garrison2014} project. In addition to encoding a diversity of accretion histories of MW-mass haloes, these simulations also have sufficient resolution such that the subhaloes hosting classical dwarfs ($\mathrm{M_{peak}} > 10^9 \mathrm{\msun}$) are not susceptible to artificial disruption \citep{vandenBosch2018a,vandenBosch2018b}.

While such dark matter-only simulations offer a number of advantages, the absence of an explicit modelling of the baryons limits their scope for the analysis of the properties of classical dwarf satellite galaxies. First, without explicit modelling of the baryonic processes, there is a degree of uncertainty as to which subhaloes host classical dwarfs \citep[Missing Satellite Problem;][]{Klypin1999,Moore1999,Garrison2017a}. Second, due to the lack of a central baryonic disk, our simulations do not account for the selective destruction of haloes on highly radial orbits passing close to the baryonic disk \citep[e.g.][]{Garrison2017b}. Furthermore, there is evidence that central baryonic disk of a MW-mass galaxy influences the orbits of dwarf satellites \citep{Gomez2017}. Hence, in this paper, we limit ourselves to study predominantly the {\it infall} properties of classical dwarfs. As large numbers of high-resolution hydrodynamical simulations become available in the future, this intuition can be extended to study the present-day properties of the surviving population of dwarf satellites of MW-mass galaxies.

The plan of this paper is as follows. In Section \ref{sec:2}, we describe the simulations of the MW-mass haloes and our methodology for identifying their massive accretions. In Section \ref{sec:3}, we examine in detail a single massive accretion. In Section \ref{sec:4}, we generalise these results to the full sample. In Section \ref{sec:data}, we compare our expectations with the data. Finally, we discuss our conclusions in Section \ref{sec:conclusions}.

\section{Numerical Methods}
\label{sec:2}
The ELVIS simulation \citep{Garrison2014} is a suite of 48 high-resolution, zoom-in simulations of MW-mass haloes (virial mass: $\mathrm{M_{vir}} = 1-3 \times 10^{12}\, \mathrm{\msun}$). Half of the ELVIS haloes reside in a paired configuration with separations and relative velocities similar to those of the MW-M31 pair, while the remainder are highly isolated haloes mass-matched to those in the pairs. The particle mass is $1.9 \times 10^5 \,\mathrm{\msun}$ and the Plummer-equivalent force softening is 140 pc. ELVIS assumes a Wilkinson Microwave Anisotropy Probe cosmology \citep[WMAP7,][$\Omega_m=0.286$, $\Omega_\Lambda=0.714$, $h=0.7$, $\sigma_8=0.82$, and $n_s=0.96$]{Larson2011}.  These high-resolution simulations reach out to 4 $\mathrm{R_{V}}$ for paired haloes, and out to 5 $\mathrm{R_{V}}$ for isolated haloes, where $\mathrm{R_{V}}$ represents the virial radius of the halo. Each simulation contains a maximum of 75 snapshots with the average temporal spacing at about 250 Myr. Further detail of the ELVIS simulations can be found in \cite{Garrison2014}. Dark matter subhaloes were identified using ROCKSTAR (Behroozi, Wechsler and Wu 2013), while merger trees were constructed using CONSISTENT-TREES algorithm (Behroozi et al. 2013). For each subhalo, its primary progenitor (main branch) is identified as the progenitor that contains the largest total mass summed from the subhalo masses over all preceding snapshots in that branch. The peak mass ($\mathrm{M_{peak}}$) of each subhalo is the maximum mass the subhalo has had over its entire main progenitor branch. Given our identified primary progenitors, we identify the main progenitor branch of each host MW-mass halo.

For each MW-mass halo, we identify all subhaloes that entered its virial radius during the course of its lifetime and label them as `associated' with the MW-mass halo. Some of these subhaloes survive to $z=0$, while others were `destroyed'. Among the surviving subhaloes at $z=0$, a large fraction of them are found within the virial radius of the host. The number of surviving subhaloes is a strong function of $\mathrm{M_{vir}}$ \citep[e.g.][]{Kravtsov2004}. Assuming a fiducial virial mass of the MW at $z=0$ as $1.33 \times 10^{12} \mathrm{\msun}$, we rescale the masses, radii and velocities of all haloes and subhaloes over cosmic time, such that $\mathrm{M}^{\prime}  =  f \, \mathrm{M}$, $\mathrm{R}^{\prime}  =  f^{1/3} \mathrm{R}$ and $\mathrm{v}^{\prime} =  f^{1/3} \mathrm{v}$, where $f= (1.33 \times 10^{12} \mathrm{\msun})/\mathrm{M_{vir}}$ and $\mathrm{M_{vir}}$ is the virial mass of the MW-mass halo at $z=0$. We identify the time of accretion or `infall time' of a subhalo ($\mathrm{t_{accretion}}$) as the first time it enters the virial radius $\mathrm{R}^{\prime}_{vir}$ of its host halo. We identify the time of `disruption' or `merger' of a subhalo ($t_{\mathrm{disruption}}$) when it coalesces with the main progenitor branch of its host halo or of another more massive halo. We restrict our attention to subhaloes with $\mathrm{M^{\prime}_{peak}} > 10^9\, \mathrm{\msun}$ which we assume are capable of hosting dwarf galaxies, some of them being classical dwarf satellites and which are not susceptible to artificial disruption \citep[e.g.][]{vandenBosch2018a, vandenBosch2018b}.

\begin{figure}
	\begin{center}
	\includegraphics[width=0.8 \columnwidth]{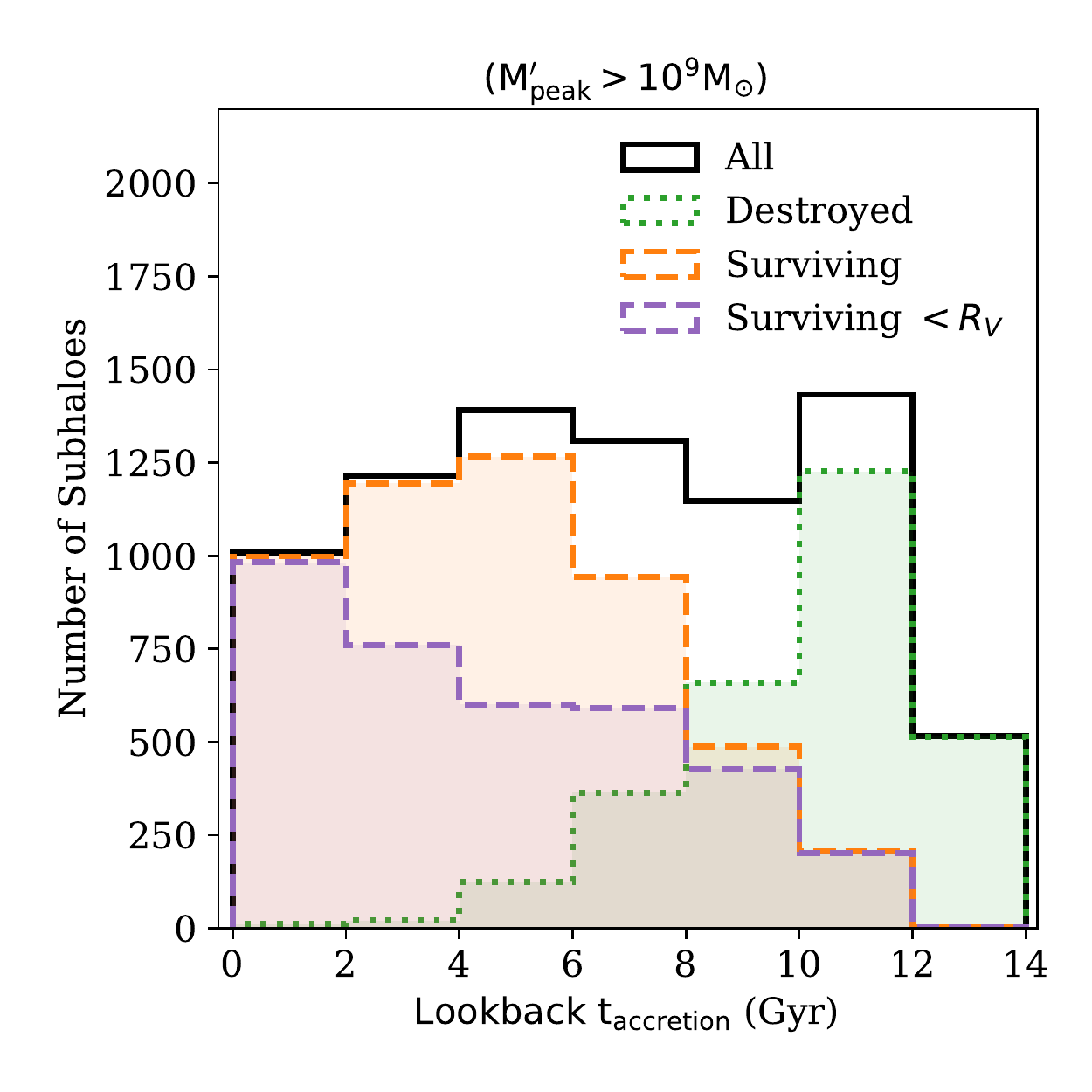}
    \caption{A histogram of the lookback time of accretion (first entry into the virial radius of the MW-mass host) of all subhaloes ($\mathrm{M^{\prime}_{peak}} > 10^9\, \mathrm{\msun}$) associated with the 48 MW-mass haloes. The lookback time of accretion of all surviving and destroyed subhaloes is also indicated. We also indicate the lookback accretion time of surviving haloes which found within the virial radius.}
    \label{fig:fig1}
	\end{center}
\end{figure}

Surviving subhaloes associated with a MW-mass host at $z=0$ distinguish themselves from destroyed subhaloes primarily in their infall time \citep[e.g.][]{Gao2004, Sales2007}. In general, surviving subhaloes were accreted recently, while destroyed subhaloes (Fig. \ref{fig:fig1}) were accreted much earlier in the history of the MW-mass halo. For the subhaloes under consideration ($\mathrm{M^{\prime}_{peak}} > 10^9\, \mathrm{\msun}$), the accretion time of the surviving subhaloes reaches as far back as 12 Gyr ago. However, the majority of surviving subhaloes were accreted more recently. Furthermore, a significant fraction of the surviving subhaloes accreted between $\sim$2 and 8 Gyrs ago are found outside the virial radius of the MW-mass halo at $z=0$ \citep[`splashback' galaxies, e.g.][]{Knebe2011, Teyssier2012}. The distribution of accretion times of the surviving subhaloes found within the virial radius bears the imprint of the absence of splashback galaxies \citep{Simpson2018,Bakels2020}. We postpone the detail examination of the formation mechanisms of splashback galaxies for a later publication. For this paper, it is sufficient to note that the galaxies found within the confines of the virial radius of the MW and M31 at $z=0$ are only a subset of the galaxies that were accreted during their lifetimes.

\begin{figure}  
	\begin{center}
    \includegraphics[width=\columnwidth]{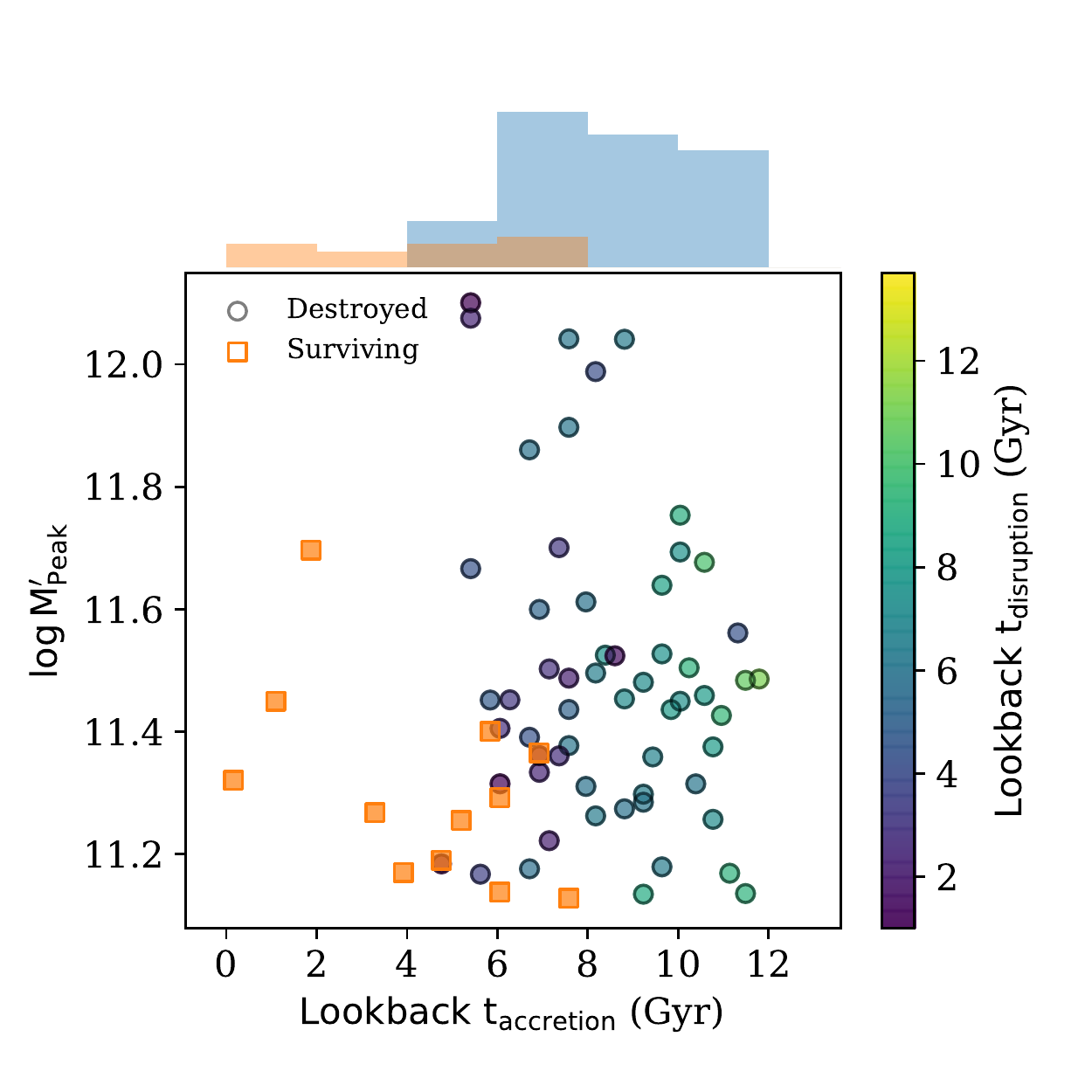}
    \caption{The peak mass ($\mathrm{M^{\prime}_{peak}}$) of the massive subhaloes accreted onto the MW-mass haloes as a function of their lookback accretion time. Orange squares show surviving massive subhaloes while circles show disrupted massive progenitors. The latter are colour-coded according to the lookback time of their disruption.}
	\label{fig:fig2}
	\end{center}
\end{figure}

\begin{figure}
\begin{center}
	\includegraphics[width=0.8\columnwidth]{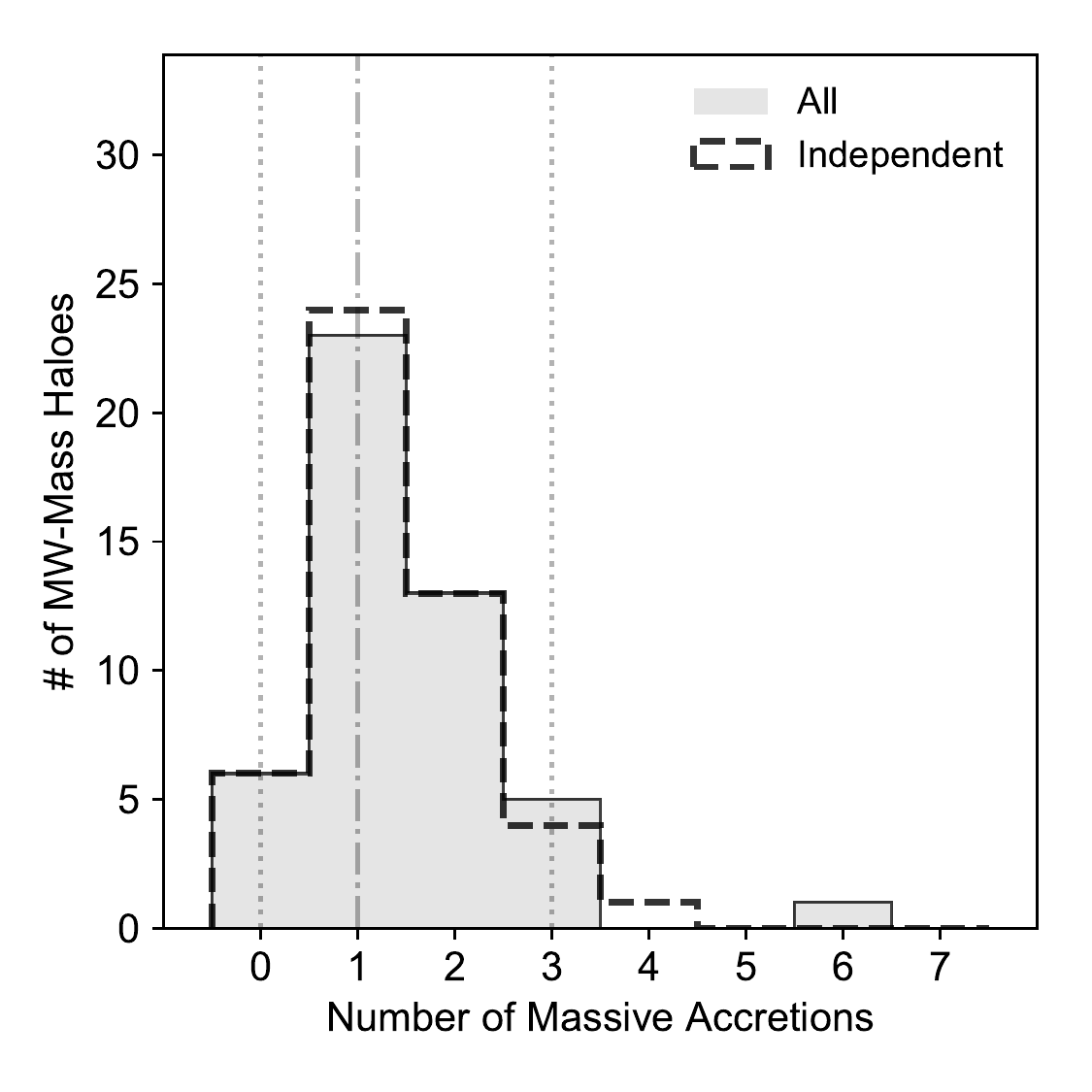}
    \caption{The frequency of MW-mass haloes which have accreted multiple massive mergers ($\mathrm{M^{\prime}_{peak}} > 1.33 \times 10^{11} \mathrm{\msun}$). The dashed vertical line indicates the median number of massive accretions, while the dotted vertical lines indicate the 10th and 90th percentiles. $\sim$12\% of the massive accreted subhaloes are accreted in pairs, i.e., one massive progenitor was already a subhalo of another massive progenitor before being accreted onto the MW-mass halo. The dashed histogram indicates the number of independent massive accreted subhaloes.} 
 \label{fig:fig3}
 \end{center}
\end{figure}

In this paper, we focus our attention on studying the massive accreted subhaloes with masses larger than 1/10 of the final halo mass ($\mathrm{M^{\prime}_{peak}} > 1.33 \times 10^{11} \mathrm{\msun}$) : those that have either merged with the main MW-mass halo or those that are still bound as subhaloes to the host at $z=0$. From the merger trees of each MW-mass halo, we identify all massive progenitors which merged with the main progenitor branch. We estimate their lookback time of accretion into the halo ($\mathrm{t_{accretion}}$) as well as the lookback time of their merger with the main progenitor branch ($\mathrm{t_{disruption}}$). We also identify all surviving massive subhaloes within the virial radius of the MW-mass halo at $z=0$, and their main progenitor branches. In the Fig. \ref{fig:fig2}, we show the peak mass {\it vs} the time of infall of the surviving/destroyed massive accreted subhaloes in our 48 MW-mass haloes. The majority of the massive subhaloes are accreted early (Fig. \ref{fig:fig2}). All disrupted massive progenitors were accreted $>5$ Gyr ago, while the majority of the surviving massive subhaloes were accreted in the last 7 Gyr. A single MW-mass halo may accrete up to a few massive progenitors in its lifetime: the median number of massive accretion events is 1, while the majority of haloes suffer between 0 and 3 accretion events (Fig. \ref{fig:fig3}), irrespective of whether the host halo is an isolated halo or one of the paired haloes. 6 out of the 48 MW-mass haloes have never accreted a massive progenitor in their lifetime. 10 out of 48 MW-mass haloes currently have a surviving massive accreted subhalo within its virial radius. Furthermore, $\sim$12\% of the massive accreted subhaloes are accreted in pairs, i.e., one massive progenitor was already a subhalo of another massive progenitor before entry into the virial radius of the MW-mass host (see Fig. \ref{fig:fig3}).

As we will see below, the accretion of a massive progenitor is coincident with the accretion of a large number of subhaloes. We choose as our primary measure a temporal criterion for exploring subhalo accretion, considering as `temporally associated' subhaloes that are accreted within 1 Gyr of the time of accretion of the massive progenitor. As a secondary criterion, and to connect with other works, we will also examine the subset of `temporal associations' that were subhalos of the massive progenitor for at least two consecutive time-steps ($\Delta \mathrm{T} \sim \mathrm{500\, Myr}$) in the simulation before entering the virial radius of the host MW-mass halo \citep{Deason2015}.  

Finally, while we restrict our attention in this work to the most massive accreted subhaloes ($\mathrm{M^{\prime}_{peak}} > 1.33 \times 10^{11} \mathrm{\msun}$; merger mass-ratio > 1:10), we note that accreted subhaloes as large $\mathrm{M^{\prime}_{peak}} \sim 8.86 \times 10^{10} \mathrm{\msun}$ (merger mass-ratio > 1:15) may also contribute a number of subhaloes to the MW-mass host. For the purpose of this paper, we will denote these massive subhaloes as `significant' accretions.

\section{Anatomy of a Massive Accretion}
\label{sec:3}
In order to develop physical intuition of a massive merger, we consider the case of a single MW-mass host (called 'iLincoln' in the ELVIS simulations), which suffered a massive accretion ($\log \mathrm{M^{\prime}_{peak}/\msun} \sim 11.39$) around 6.5 Gyr ago. In addition to this massive accretion, this MW-mass halo also accreted two additional significant progenitors of masses $\log\,\mathrm{M^{\prime}_{peak}/\msun}\sim10.9$  and $\log\,\mathrm{M^{\prime}_{peak}/\msun}\sim10.73$  around 10.2 and 1.1 Gyr ago respectively. In particular, we focus on the most massive accretion $\sim$6.5 Gyr ago and the subhaloes which were accreted along it.

\subsection{Accretion of a large number of subhaloes}
\begin{figure}
\begin{center}
	\includegraphics[width=0.9\columnwidth]{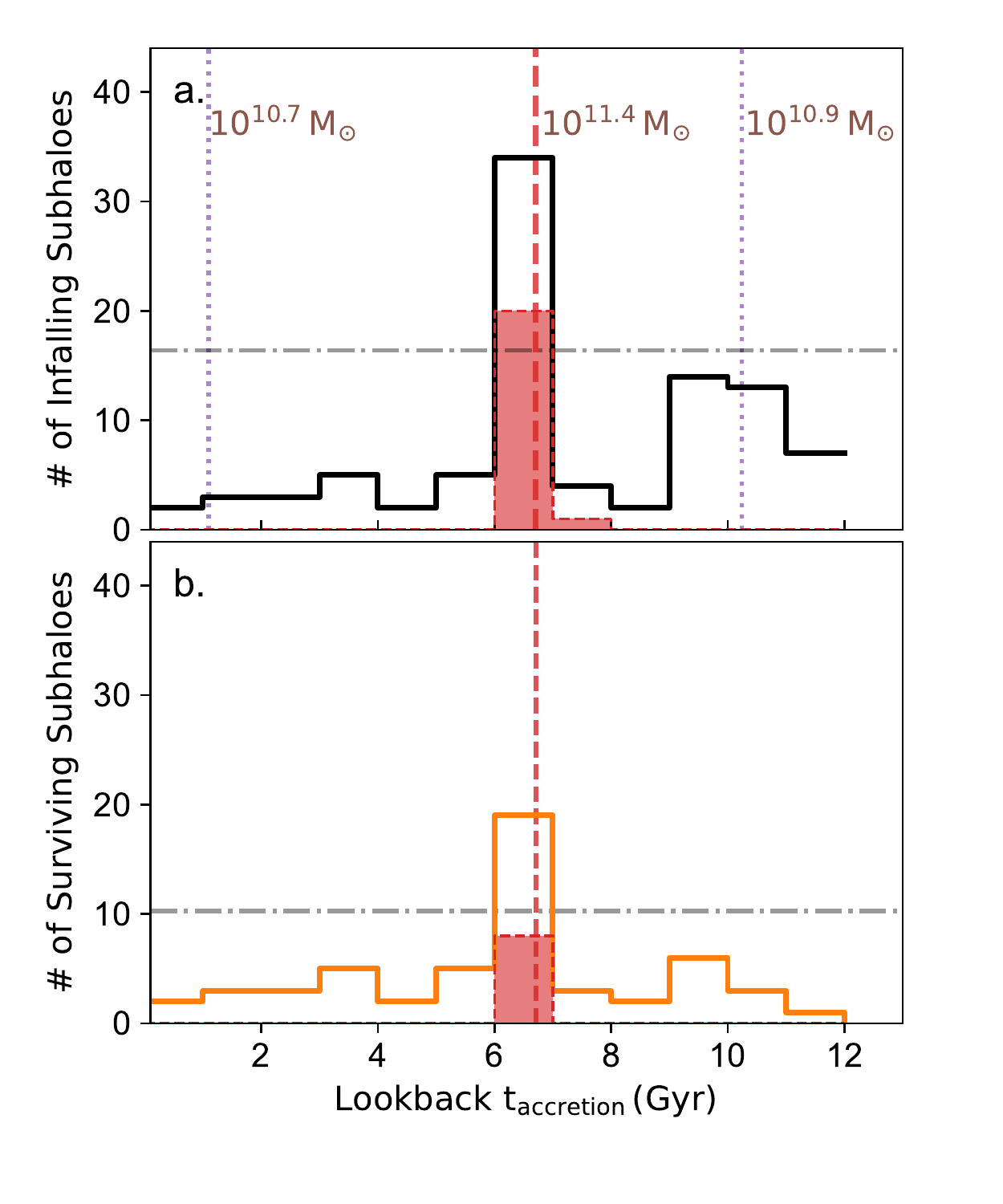}
    \caption{Top: The number of all infalling subhaloes ($\mathrm{M^{\prime}_{peak}} >  10^9 \mathrm{\msun}$; found within and outside the virial radius) as a function of their lookback time of accretion of a single MW-mass halo `iLincoln'. The dashed vertical line indicates the accretion time of a single massive progenitor of mass $\log(\mathrm{M^{\prime}_{peak}})\sim11.39$ around 6.5 Gyr ago. Two additional significant progenitors, $\log(\mathrm{M^{\prime}_{peak}})\sim10.9$ and $\log(\mathrm{M^{\prime}_{peak}})\sim10.7$ were accreted 10.2 and 1.1 Gyr ago respectively. The horizontal line indicates the 3 $\sigma$ standard deviation from the average infall rate of subhaloes per Gyr. The shaded histogram indicates the assigned subhaloes of the massive progenitor according to the criterion of Deason et al. 2016. Bottom: The number of `surviving' subhaloes at $\mathrm{z=0}$ as a function of their accretion time. Similarly, the horizontal line indicates the 3 $\sigma$ standard deviation from the average infall rate of surviving subhaloes at $\mathrm{z=0}$ per Gyr.}
    \label{fig:fig4}
\end{center}
\end{figure}

Along with the massive progenitor, a large number of subhaloes are accreted onto the host galaxy, i.e., enter its virial radius. In Fig. \ref{fig:fig4}, we plot the number of infalling subhaloes ($\mathrm{M^{\prime}_{peak}} >  10^9 \mathrm{\msun}$) as a function of their time of accretion. We also indicate in the lower panel the number of infalling subhaloes which survive until $\mathrm{z=0}$. We find that the infall time of subhaloes is strongly clustered around the massive accretion $\sim$6.5 Gyr ago. We also find similar but smaller clusterings in infall time around the other significant accretion $\sim$10.2 Gyr ago, but find no sign of clustering for the infall at $\sim$1.1 Gyr ago.  The rate of infall of subhaloes accreted along with the massive progenitor is significantly higher than the average infall rate of subhaloes per Gyr. The clustering in infall time is also visible among the surviving subhaloes at $\mathrm{z=0}$. This suggests that the infall time distribution of observed satellites encodes information about massive accretions of a MW-mass galaxy.

The number of subhaloes accreted also appears to be in excess of the number of subhaloes `physically' associated with the massive progenitor. It is common to use the \cite{Deason2015} criterion for a physical association (Fig. \ref{fig:fig4}), which considers only those subhaloes which have been spatially associated with the massive progenitor for at least 500 Myr (2 snapshots in ELVIS) before its infall into the virial radius of the MW-mass host. For this particular halo, only 54\% of the subhaloes accreted within 1 Gyr of the time of accretion of the massive progenitor satisfy the \cite{Deason2015} criterion. This suggests that a temporal association of subhalos with a massive accretion might  be a more productive way to think about subhalo accretion, as this would account for the excess subhalos accreted at the same time as the massive progenitor. Given such a temporal criterion (accretion within 1Gyr of the massive progenitor), we find that nearly 35\% of the subhaloes are accreted along with the massive progenitor (Fig. \ref{fig:fig4}).

\begin{figure*}
\begin{center}
	\includegraphics[width=0.9 \textwidth]{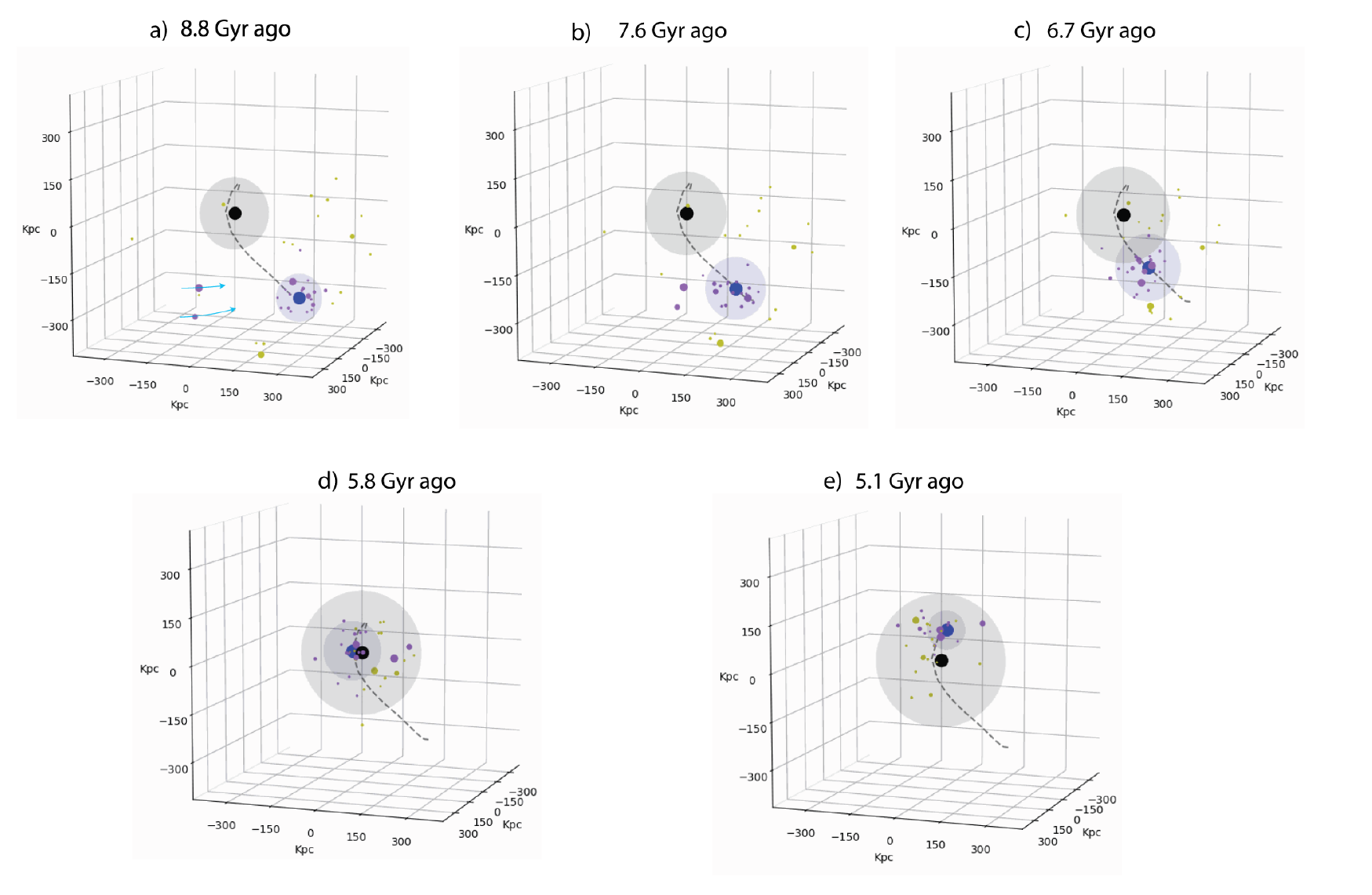}
    \caption{{\it The anatomy of a massive accretion}: The panels show the positions of the subhaloes of `iLincoln' accreted within $+/-$ 1 Gyr of the massive progenitor in physical galactocentric coordinates at various snapshots (pre-accretion, during-accretion and post accretion). The massive progenitor was accreted 6.5 Gyr ago. The panels represent snapshots at a) 8.8, b) 7.6, c) 6.7, d) 5.8 and e) 5.1 Gyr ago. In panel d), the massive progenitor is at the first pericenter. The central black circle shows the position of the main MW-mass halo, while the blue circle shows the position of the massive progenitor. The spheres around the black and the blue circles represent the virial radii of the main MW-mass halo and the massive progenitor. The black dashed-line shows the path of the massive progenitor. The violet circles represent haloes which are assigned as subhaloes of the massive progenitor according to the criterion of Deason et al. 2015. The yellow circles represent haloes which are not assigned as subhaloes to the massive progenitor but which were accreted within $+/-$ 1 Gyr of the massive progenitor. The size of all haloes and subhaloes are scaled according to their peak mass, except the central MW-mass halo and the massive accreted progenitor due to aesthetical reasons. In panel a) we highlight the velocity vector of two haloes which will later be associated with the massive progenitor according to the Deason et al. criterion and which have different orbits compared to the massive progenitor.}
    \label{fig:fig5}
\end{center}
\end{figure*}

\begin{figure}
\begin{center}
	\includegraphics[trim= 1in 0.2in 1in 0.2in, width=0.8\columnwidth]{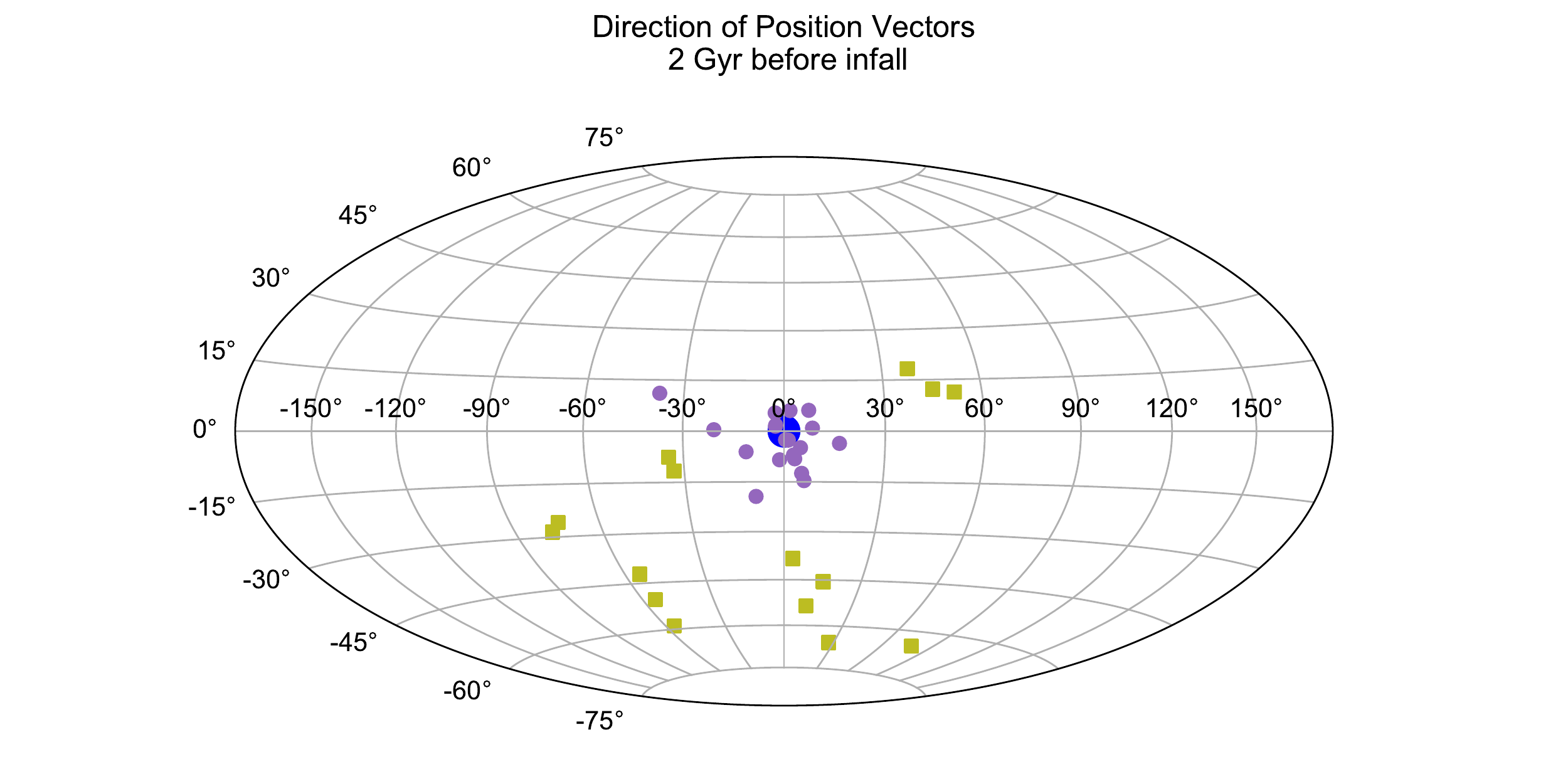}
    \caption{{\it Pre-Accretion Phase}: The difference in angles between the position vector of the massive progenitor and its co-accreted subhaloes 2 Gyr before the accretion of the massive progenitor onto the MW-mass halo `iLincoln'. All subhaloes are accreted within $+/-$ 1 Gyr of the massive progenitor. The blue-circle shows the position of the massive progenitor. The violet circles represent haloes which are assigned as subhaloes of the massive progenitor according to the criterion of Deason et al. 2015, while the yellowish-filled squares are not assigned as subhaloes of the massive progenitor, but fall in at the same time.}
    \label{fig:fig6}
\end{center}
\end{figure}

A visual examination of the geometry of the accretion of the massive progenitor illustrates the origin of the subhaloes accreted along with the massive progenitor. In Fig. \ref{fig:fig5}, we show a series of 5 snapshots depicting the progress of the ongoing accretion event. In particular, the snapshots highlight the pre-accretion ($\sim$2.5 Gyr before), entry and post-accretion ($\sim$1.5 Gyr after) stages of the infall event. Surprisingly, in the pre-accretion phase, the subhaloes appear be attracted towards the MW-mass host from a large solid angle in the sky. Fig. \ref{fig:fig6} shows the angular separations of the subhaloes accreted along with the massive progenitor $\sim$ 2 Gyr before the time of its accretion. Subhaloes tagged through the Deason et al. criterion have separations less than 30 degrees from the massive progenitor. However, some subhaloes which will eventually infall along with the massive progenitor are initially as far as 60 degrees away from it on the sky. As the massive progenitor approaches the MW-mass host, these distant subhaloes appear to be attracted towards the massive progenitor and further onto the MW-mass host, and hints to the fact that the accretion of a massive merger leads to a clustering in the infall time of subhaloes.

\begin{figure}
\begin{center}
	\includegraphics[trim= 0in 1.3in 0in 0.0in, width=0.8\columnwidth]{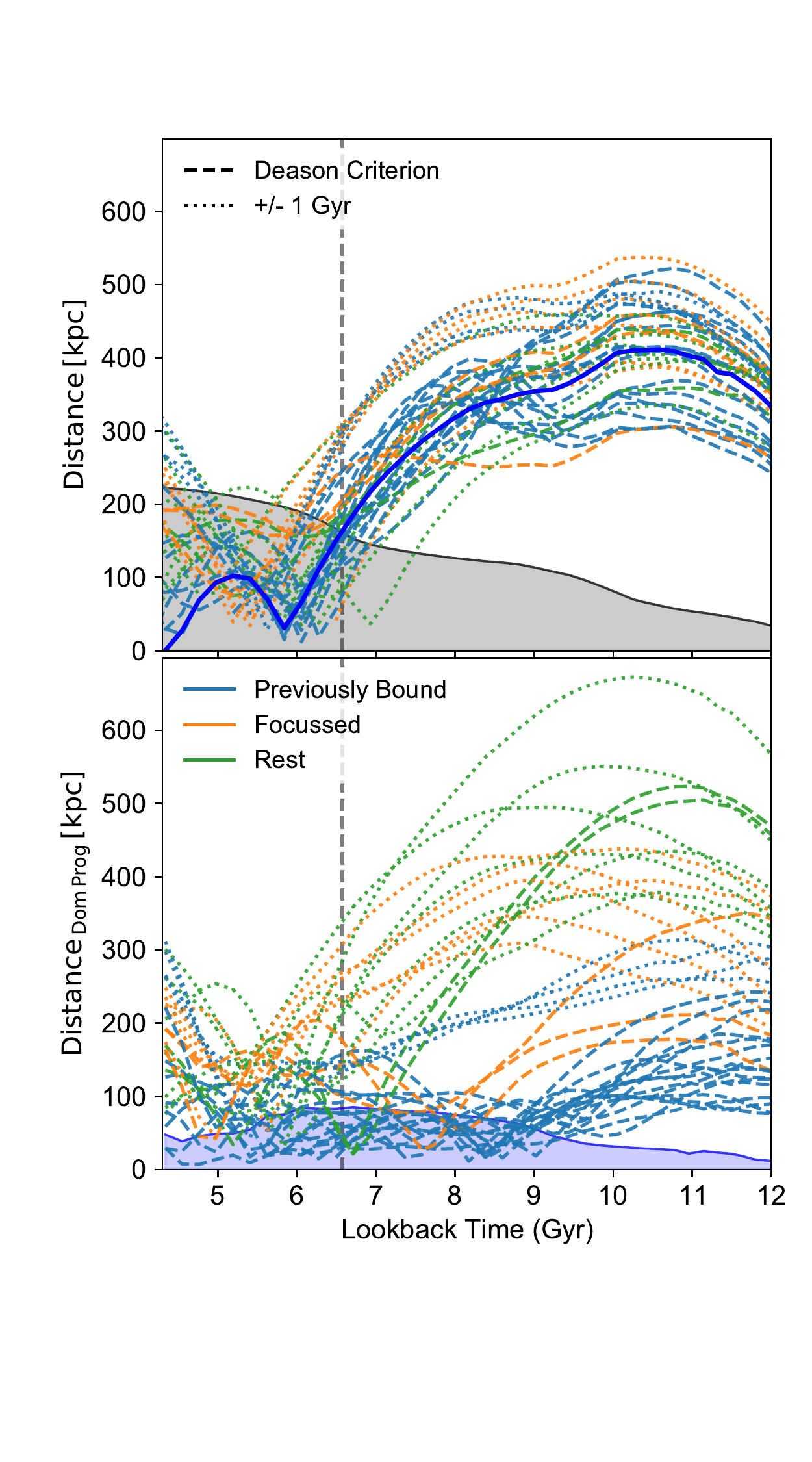}
	\caption{The top panel represents the physical galactocentric distances as a function of lookback time of the subhaloes of of `iLincoln' accreted within $+/-$ 1 Gyr of the massive progenitor, while the lower panel represents their physical distances from the massive progenitor as a function of lookback time. The solid blue line in the top panel represents the massive progenitor. The black (top panel) and blue (lower panel) solid lines enclosing the shaded regions represent the virial radius of the MW-mass host and the massive progenitor respectively. In both panels, we represent the subhaloes selected along with the Deason et al. criterion with dashed lines, while the remaining subhaloes accreted within 1 Gyr of the massive progenitor are represented with dotted lines. Infalling subhaloes are further classified according into three groups: a) blue indicates subhaloes which are bound to the massive progenitor. b) Orange represents subhaloes which have been gravitationally focused by the massive progenitor but are not bound to the massive progenitor. c) Green represents subhaloes that were neither bound nor were gravitationally-focused but which are accreted within 1 Gyr of the massive progenitor.}
	\label{fig:fig7}
\end{center}
\end{figure}

An insight into the origin of this clustering of the infall time of subhaloes can be gleaned by considering their individual paths. In Fig.\  \ref{fig:fig7}, we attempt to visualise the paths of all subhaloes accreted with 1 Gyr of the massive progenitor, i.e., we plot their physical distances from the central MW-mass host and the massive progenitor as a function of lookback time. Such an exercise allows one to note that the beginning of the clustering of the subhaloes around the massive progenitor coincides when it is most distant from the MW-mass host. Far away from the MW-mass host, the most immediate potential that these subhaloes experience is that of the massive progenitor. This suggests that the ability of the massive accreted progenitor to gravitationally focus surrounding subhaloes onto itself is maximum when it is most distant from the MW-mass host.

We explore this issue further using simple dynamical considerations. Subhaloes at distances less than $\mathrm{b_{f}}=2 \mathrm{G\,M_{Dom}}/\mathrm{v^2_{r}}$ from the massive progenitor, where $\mathrm{v_{r}}$ is the relative velocity between them, will experience a gravitational tug onto the massive progenitor \citep[e.g.][]{2003gmbp.book}. While subhaloes experience competing gravitational forces from various perturbers,  following \cite{2003gmbp.book} we consider that a subhalo is `gravitationally-focussed' by the massive progenitor if its distance of closest approach to the massive progenitor prior to MW-mass host infall is less than $\mathrm{b_{f}}$. For practical purposes, we identify those subhaloes that were bound to the massive progenitor for at least 1 Gyr in the last 5 Gyr as `previously bound', while classifying all other gravitationally-focused subhaloes as `focused'. Finally, we include in a third group (labelled the `rest') the remaining subhaloes which are too distant from the massive progenitor to have been gravitationally focused. It is worth reiterating that in our classification scheme, subhaloes that were `previously bound' have also been gravitationally-focused onto the massive progenitor.

\begin{figure}
\begin{center}
	\includegraphics[width=0.8\columnwidth]{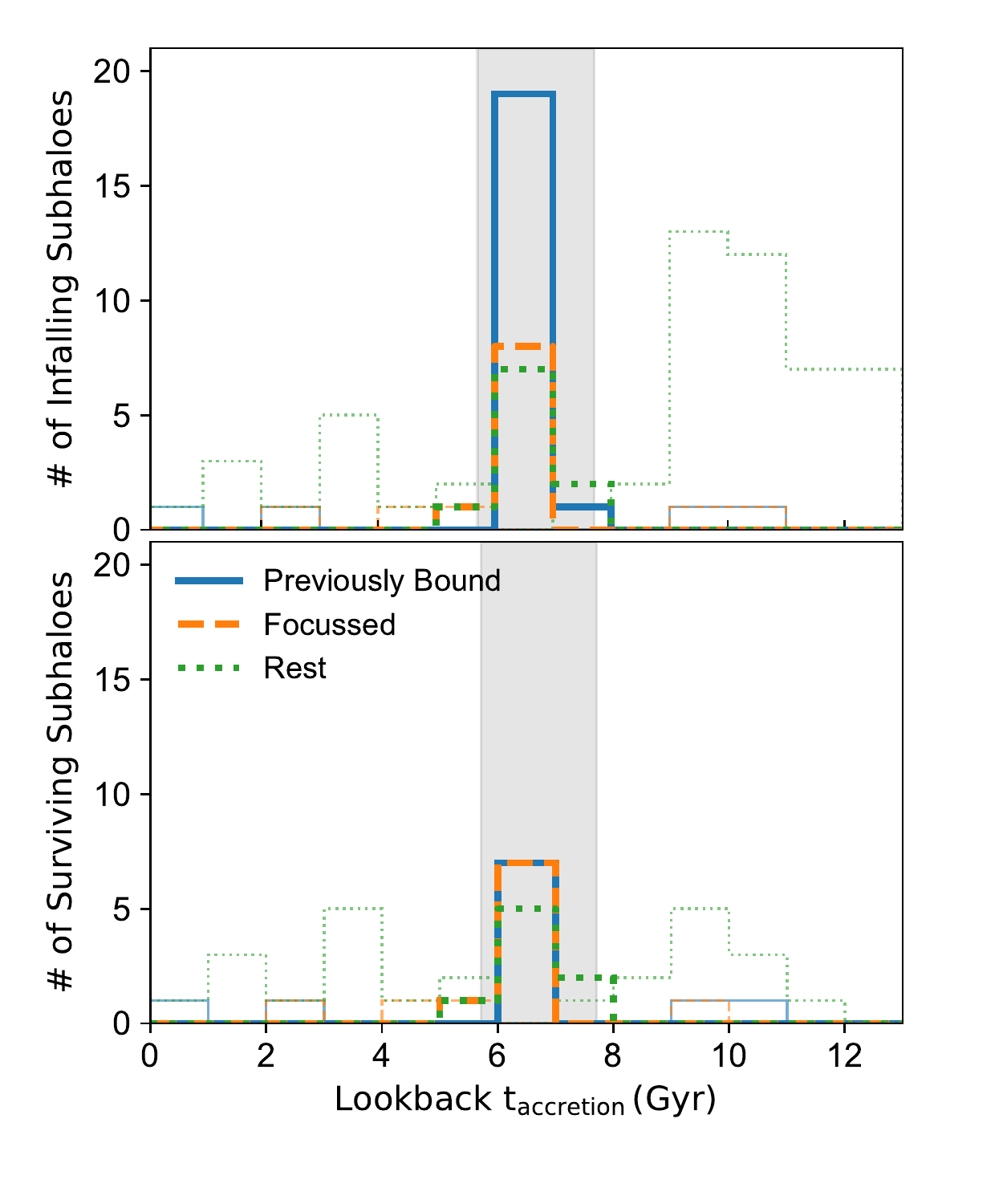}
	\caption{Top: The number of all infalling subhaloes as a function of their lookback time of accretion of a single MW-mass halo `iLincoln' divided now into 3 categories. The blue histogram represents the subhaloes which were previously bound to the massive progenitor. The orange histogram represents subhaloes which were gravitationally focused by the massive progenitor but were never bound to the massive progenitor. The green histogram represents subhaloes that were neither bound nor were gravitationally-focused but which are accreted along with the massive progenitor. The histograms are also divided into two phases: within 1 Gyr of the accretion of the massive progenitor, and outside this period. Bottom: Same as above, but illustrating only the surviving subhaloes of `iLincoln'.}
	\label{fig:fig8}
\end{center}
\end{figure}

With such a classification scheme, the tracks of subhaloes accreted within 1 Gyr of the massive progenitor in Fig. \ref{fig:fig7} differentiate into three broad classes. The Deason et al. criterion does a fairly good job in identifying most of the subhaloes that were `previously bound' to the massive progenitor. However, since it is a proximity-based criterion, it misses a few subhaloes which are loosely bound to the massive progenitor, while including 2 subhaloes which were accreted by the massive progenitor but for whom the MW-mass host's gravity was always dominant. Many of the subhaloes that were `previously bound' have completed a number of pericentric passages close to the massive progenitor, potentially affecting their star formation properties. From Fig. \ref{fig:fig8}, a large fraction of the subhaloes accreted with 1 Gyr of the massive progenitor were `previously bound' to it, and can be considered as subhaloes belonging to the massive progenitor. Furthermore, there also appears to be a substantial accretion of subhaloes along with the massive progenitor that do not appear to be gravitationally-focused by it. We postulate that this `correlated accretion' may be associated with the rapid growth of the central MW-mass halo as it accretes the massive progenitor. While a few of the subhaloes that have been gravitationally-focused by the massive progenitor are accreted at later or earlier times, the majority of them are accreted within 1 Gyr of the massive progenitor. Finally, in relation to the other two groups, a large fraction of the `previously bound' subhaloes are eventually destroyed and do not survive till the present day.

\begin{figure}
\begin{center}
	\includegraphics[width=0.8\columnwidth]{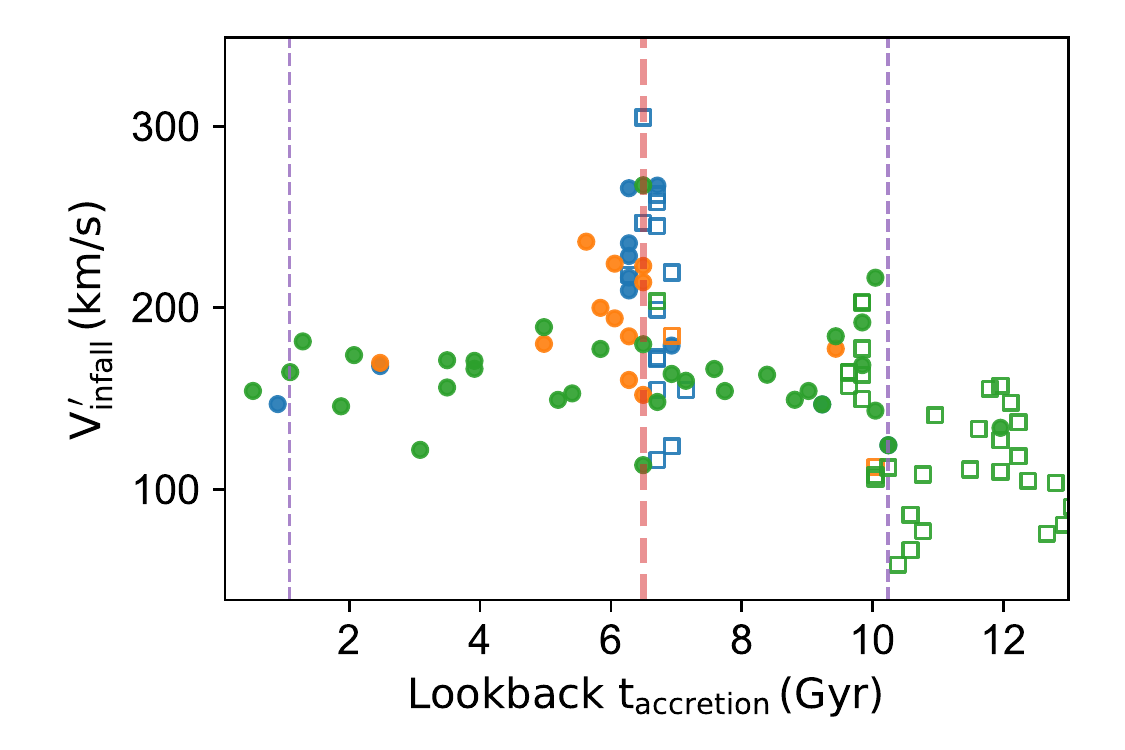}
    \caption{The infall velocity of accreted subhaloes of `iLincoln' when they cross the virial radius as a function of their accretion time. The subhaloes follow the same colour scheme as in Fig. 8: Blue represents subhaloes `previously bound' to the massive progenitor. Orange represents subhaloes which were gravitationally focused by the massive progenitor but were never bound to the massive progenitor, while green represents the rest of the subhaloes. Subhaloes which do not survive to the present day are shown as empty squares, while subhaloes which survive till the present day are shown in filled circles. }
	\label{fig:fig9}
\end{center}
\end{figure}

The differences in the origins of the subhaloes accreted along with a massive progenitor leads to differences in their infall velocities. In Fig.\ \ref{fig:fig9}, we plot the infall velocity of the individual subhaloes as they cross the virial radius of the MW-mass host as a function of their infall time. Subhaloes accreted along with the massive progenitor 6.5 Gyr ago have a large spread in their infall velocities, with some subhaloes having infall velocities larger than 200 km/s. In particular, subhaloes previously bound to the massive progenitor have significantly higher infall velocities than those subhaloes which were not gravitationally-focused by it. Another similar spread in infall velocities also coincides with the accretion of a significant progenitor accreted 10.2 Gyr ago. We hypothesise that the large spread in infall velocities of subhaloes accreted along with a massive progenitor may lead them to interact with the circumgalactic medium of the MW-host in different ways, possibly resulting in important differences in their star formation properties.

While there are many merits in distinguishing subhaloes according to their infall orbits, it remains a difficult if not impossible task to observationally classify the infall orbits of surviving dwarf galaxies based on their present-day phase-space properties. Since we wish to constrain the infall time of dwarf galaxies (within a certain mass range) from their star formation shutdown times (see Section \ref{sec:data}), we choose for the remainder of the paper to retain the 1 Gyr temporal association criterion, while making comparisons with the Deason et al. subhalo tagging criterion.

\begin{figure}
\begin{center}
    \includegraphics[trim= 1in 0.2in 1in 0.2in, width=0.8 \columnwidth]{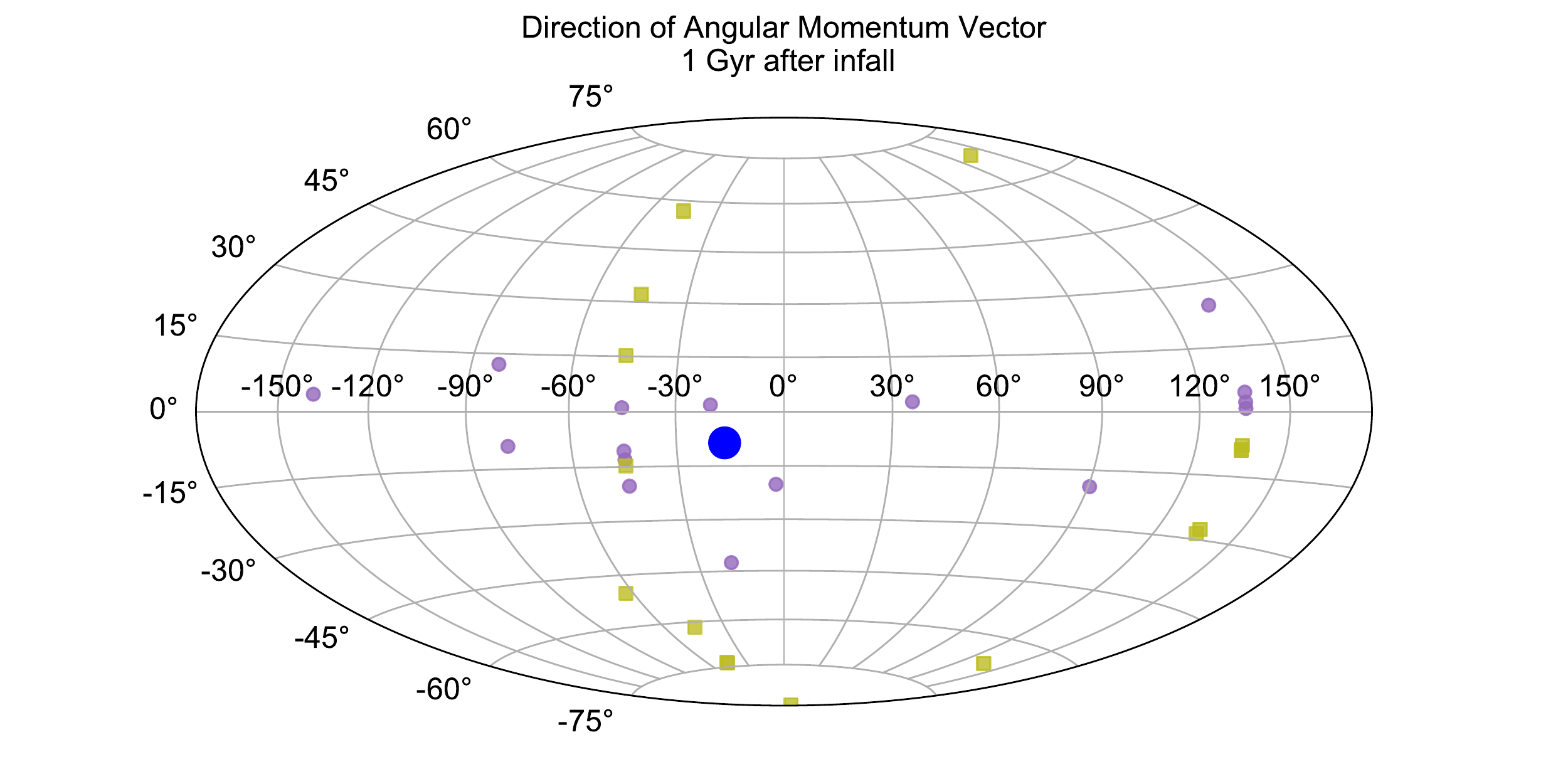}
    \caption{{\it Post-Accretion Phase}: The direction of the angular momentum vector of the subhaloes accreted along  with the massive progenitor 1 Gyr after the latter enters the virial halo of the host `iLincoln'. The blue circle signifies the massive progenitor. Violet filled circles are classified as subhaloes of the massive progenitor according to the criterion of Deason et al. (2015), while the yellowish-filled squares are not assigned subhaloes of the massive progenitor.}
    \label{fig:fig10}
\end{center}
\end{figure}

The group infall of the subhaloes along with the massive progenitor results in a diversity of orbits in the post-accretion phase. Fig. \ref{fig:fig10} shows the direction of the orbital poles of the subhaloes accreted along with the massive progenitor on the sky $\sim$ 1 Gyr after the time of accretion. There is a large scatter in their orbital poles and are not restricted to a single plane. A few of the accreted subhaloes possess an opposite direction of the angular momentum vector to that of the massive progenitor, including associated subhaloes according to the stricter criterion of \cite{Deason2015}. A number of reasons contribute to the large scatter in the post-accretion infall orbits. First, there is a systematic difference in the orbital poles of the massive progenitor in the pre-accretion and post-accretion phases. Second, since the radius of accreted group is comparable to the impact parameter and the size of the MW-host (1:4 merger), infalling subhaloes pass around the massive primary in a variety of ways, giving rise to a diversity of orbital poles after first passage. Indeed, some subhaloes may pass around the opposite side of the primary as the most massive accreted subhalo, giving an opposite sense of angular momentum.

\subsection{Destruction of subhaloes}
\label{sec:3destruct}
\begin{figure}
\begin{center}
	\includegraphics[width=0.8\columnwidth]{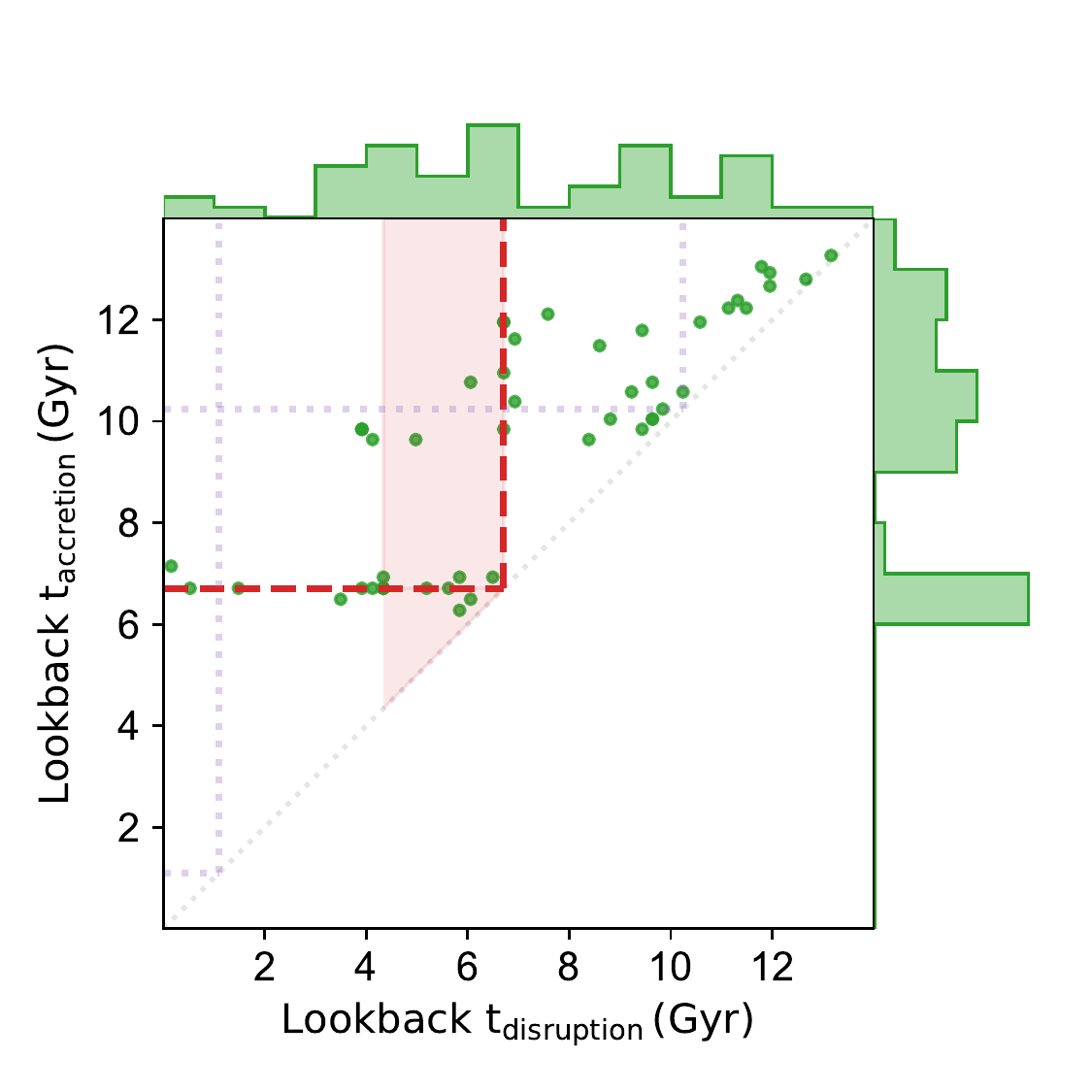}
    \caption{The time of accretion {\it vs} the time of destruction of all accreted `destroyed' subhaloes of the MW-mass host `iLincoln'. The shaded region indicates the time the massive accreted progenitor spends within the virial radius of the MW-mass halo before being eventually destroyed.}
    \label{fig:fig11}
\end{center}
\end{figure}

The accretion of a massive progenitor also precipitates the destruction of a large number of subhaloes. From Fig. \ref{fig:fig9}, it is clear that some subhaloes that are accreted along with the massive progenitor do not survive to the present day. In particular, those subhaloes which enter the virial radius of the MW-mass host slightly ahead of the massive progenitor appear to be preferentially destroyed. In Fig. \ref{fig:fig11}, we plot the time of accretion {\it vs} the time of merger (or disruption) of all the accreted subhaloes which do not survive until $z=0$. The time of disruption indicates the epoch when the subhalo branch merges with the main progenitor branch or another more massive subhalo. Once accreted, a subhalo can be destroyed immediately or after a considerable time delay. We also indicate in Fig. \ref{fig:fig11} the time that the massive progenitor spends within the virial radius of the MW-mass halo before it eventually merges. A number of subhaloes accreted along with the massive accreted progenitor are eventually destroyed. In particular, a sizeable fraction of these accreted subhaloes are destroyed while the massive progenitor is orbiting within the virial radius of the MW-mass halo, and when it begins to dramatically and suddenly influence the overall potential of the system and in particular, the orbits of nearby subhaloes within its sphere of influence. A few existing subhaloes with stable orbits around the central MW-mass host are also disturbed by the entry of the massive progenitor, resulting in the destruction of a number of previously-accreted subhaloes. Simple analytic considerations of the sphere of influence suggest that the ability of the massive progenitor ($\mathrm{R_{v}}\sim100$ kpc) to influence the orbits of subhaloes is at its maximum when it is at a galactocentric distance of less than 100 kpc. We find no difference between the shape of the mass function of the subhaloes destroyed due to the accretion of a massive progenitor from that of the rest of the subhaloes.

\begin{figure}
\begin{center}
	\includegraphics[width=0.8\columnwidth]{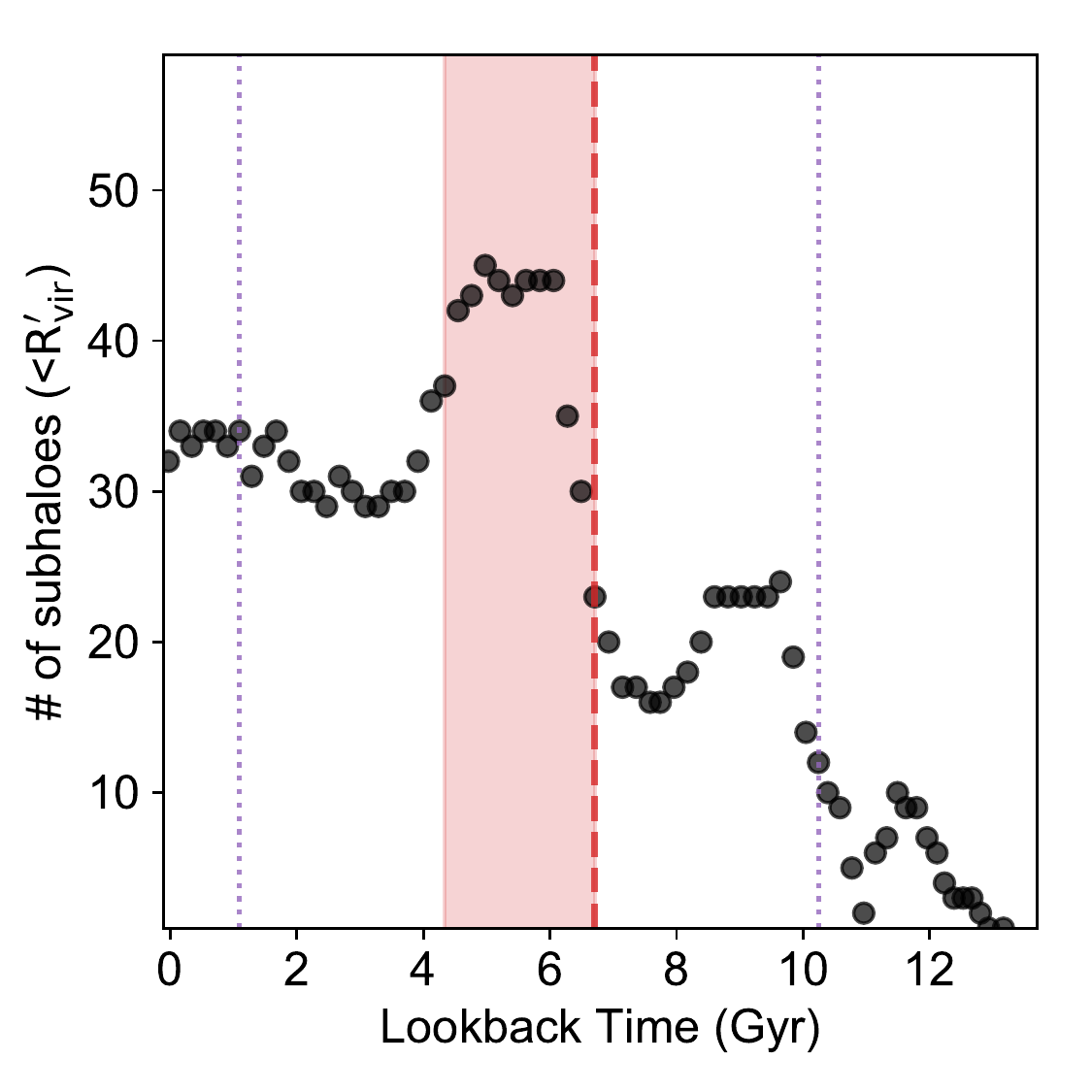}	
	\caption{The number of subhaloes of the MW-mass halo `iLincoln' out to the virial radius ($\mathrm{R^{\prime}_{vir}}$) as a function of a function of lookback time. The red dashed vertical line indicates the time of accretion of the massive progenitor. The violet dotted vertical lines indicates the time of accretion of the other two significant progenitors.}
    \label{fig:fig12}
\end{center}
\end{figure}

The accretion and destruction of subhaloes along with a massive progenitor results in a temporary excess of dwarf galaxies with the virial radius of the MW-mass host. Along with a massive accretion, the number of subhaloes of a MW-mass host within a certain galactocentric radius first increases rapidly, and subsequently decreases (see Fig. \ref{fig:fig12}). The decrease in the number of subhaloes is caused by i) the destruction of some subhaloes and ii) the exit of some subhaloes with large apocenters beyond the virial radius of the host. Furthermore, a number of subhaloes accreted along with the massive progenitor with large apocenters may renter the virial radius at later times.

\begin{figure}
\begin{center}
	\includegraphics[width=0.8\columnwidth]{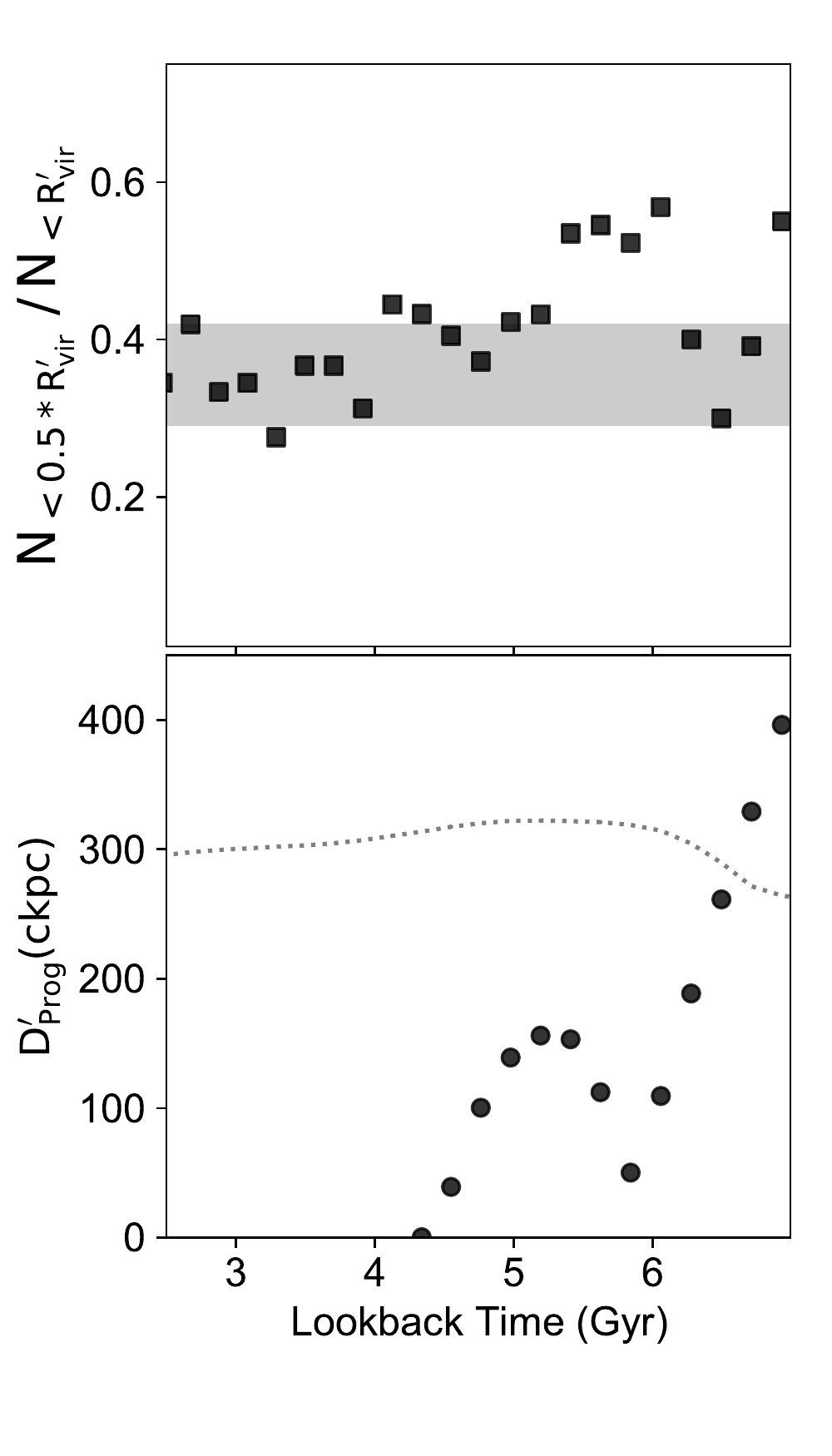}
    \caption{Top: The fraction of subhaloes found within 0.5 $\times$ $\mathrm{R^{\prime}_{vir}}$ of the MW-mass halo as a function of lookback time. The horizontal shaded region is the average stable value of the fraction. Bottom: The radial co-moving distance of the massive progenitor as a function of lookback time. The dotted line represent the virial radius of the MW-mass halo as a function of lookback time. The accretion of a massive progenitor redistributes the radial distribution of subhaloes.}
    \label{fig:fig14}
\end{center}
\end{figure}

The accretion and destruction of subhaloes along with a massive accretion also temporarily affects the radial profile of subhaloes and satellites. To demonstrate this, we characterise the radial profile of a subhaloes through a concentration parameter defined as $\mathrm{N_{0.5\,R^{\prime}_{vir}}/N_{R^{\prime}_{vir}}}$. In Fig. \ref{fig:fig14}, we see that the concentration of the cumulative radial profile of the subhaloes changes according to the position of the massive progenitor along its orbit. When the massive progenitor is close to the first pericenter, the cumulative radial profile becomes more centrally concentrated, with an excess of satellites at small galacto-centric distances $<150$ kpc. During various phases of the orbit of the massive progenitor, the radial profile of the subhaloes can be less centrally concentrated than the average radial profile. These changes in the radial profile are relatively short-lived and last less than 600 Myr.  While other physical processes may affect the distribution of subhaloes (a study of which is beyond the scope of the paper), the accretion of a massive progenitor has a strong {\it temporary} effect on the radial profile of subhaloes and satellites.

\section{Massive Accretions in Milky Way haloes}
\label{sec:4}
We now turn our attention to how these results generalise to a sample of MW-mass hosts, and how massive accretions affect the properties of the subhalo population.

\subsection{Accretion and Destruction of subhaloes}
\begin{figure}
	\begin{center}
    \includegraphics[width=0.8 \columnwidth]{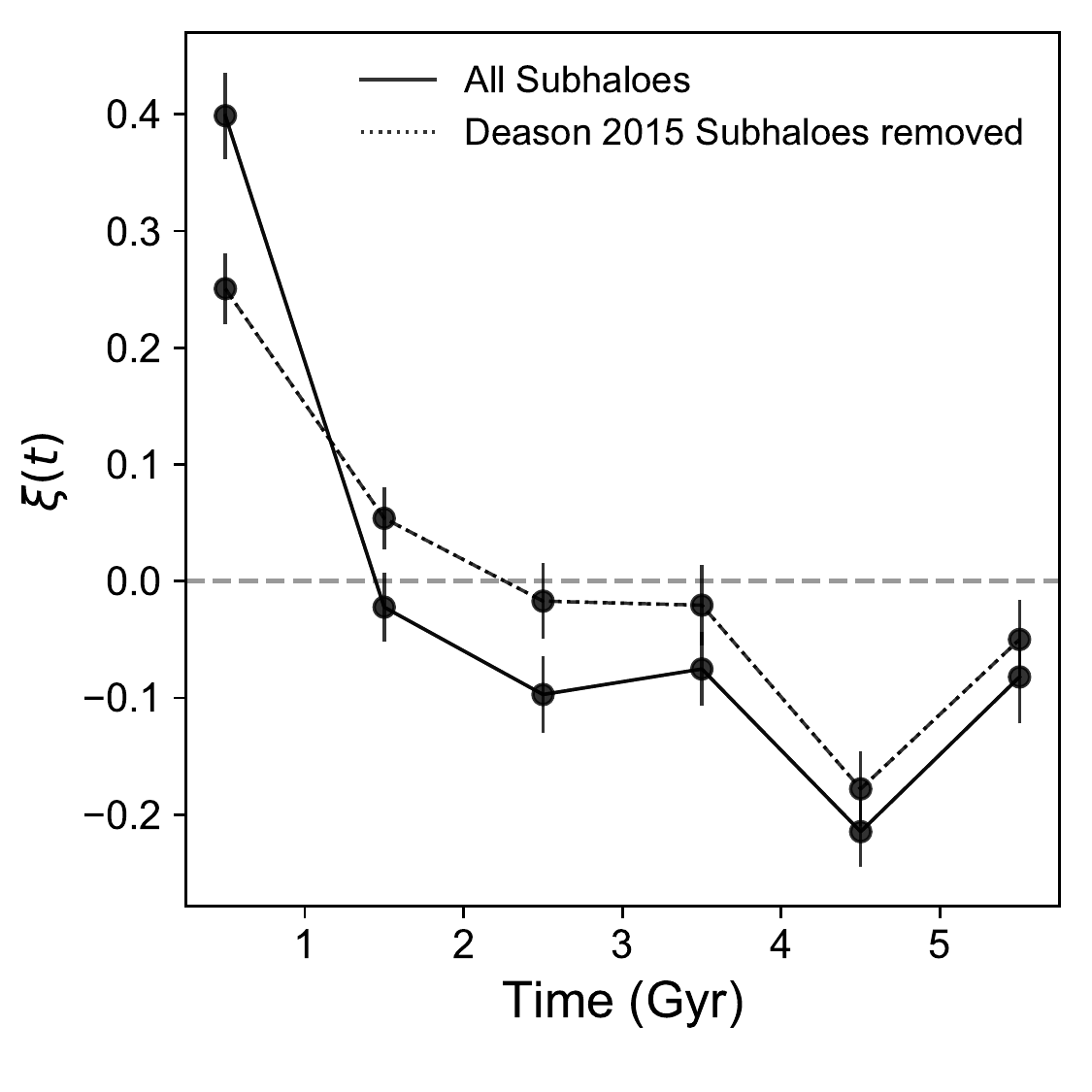}
    \caption{Accretion of subhaloes by massive progenitors: Cross-correlation function of the infall time of massive progenitors with the infall time of all subhaloes. Also shown is the same cross-correlation function omitting the subhaloes tagged to the massive progenitor according to the Deason et al. criterion.}
    \label{fig:fig15}
	\end{center}
\end{figure}

First, we attempt to quantify the accretion and destruction of subhaloes along with massive progenitors in MW-mass haloes using a cross-correlation function. To measure the accretion of subhaloes, we use a cross-correlation function between the time of accretion of the massive progenitors and the time of accretion of all subhaloes. To construct such a cross-correlation function, we first generate a random sample of the infall time of subhaloes using the distribution obtained in Fig. \ref{fig:fig1}. We then use the estimator
\begin{equation}
	\xi(t) = \frac{N_{R}} {N_{S}} \frac{DD(t)}{DR(t)} - 1
\end{equation}
where $N_{R}$ is the number of points in our random catalogue, $N_{S}$ is the number of subhaloes, $DD(t)$ is the count of pairs (subhaloes-massive progenitors) accreted within time $t$ and $t +\delta t$ and $DR(t)$ is the count of pairs (random-massive progenitors).  The resulting cross-correlation function is presented in Fig. \ref{fig:fig15}. We find that the infall of subhaloes in a MW-mass galaxy is clearly clustered around the accretion time of the massive progenitors. The decrease in the cross-correlation function suggests that the time frame within which subhaloes are accreted along with the massive progenitor is +/- 1 Gyr, reinforcing the temporalcriterion we have used to select subhaloes accreted along with the massive progenitor. Similar results are obtained by cross-correlating the infall time of just the surviving subhaloes at $z=0$ with the infall time of the massive progenitors. Furthermore, even after omitting the subhaloes tagged as belonging to the massive progenitor according to the Deason et al. criterion, the cross-correlation still persists (Fig. \ref{fig:fig15}), suggesting that the number of subhaloes accreted along with the massive progenitor is in excess of what is captured by the Deason et al. criterion.

\begin{figure}
	\begin{center}
    \includegraphics[width=0.8 \columnwidth]{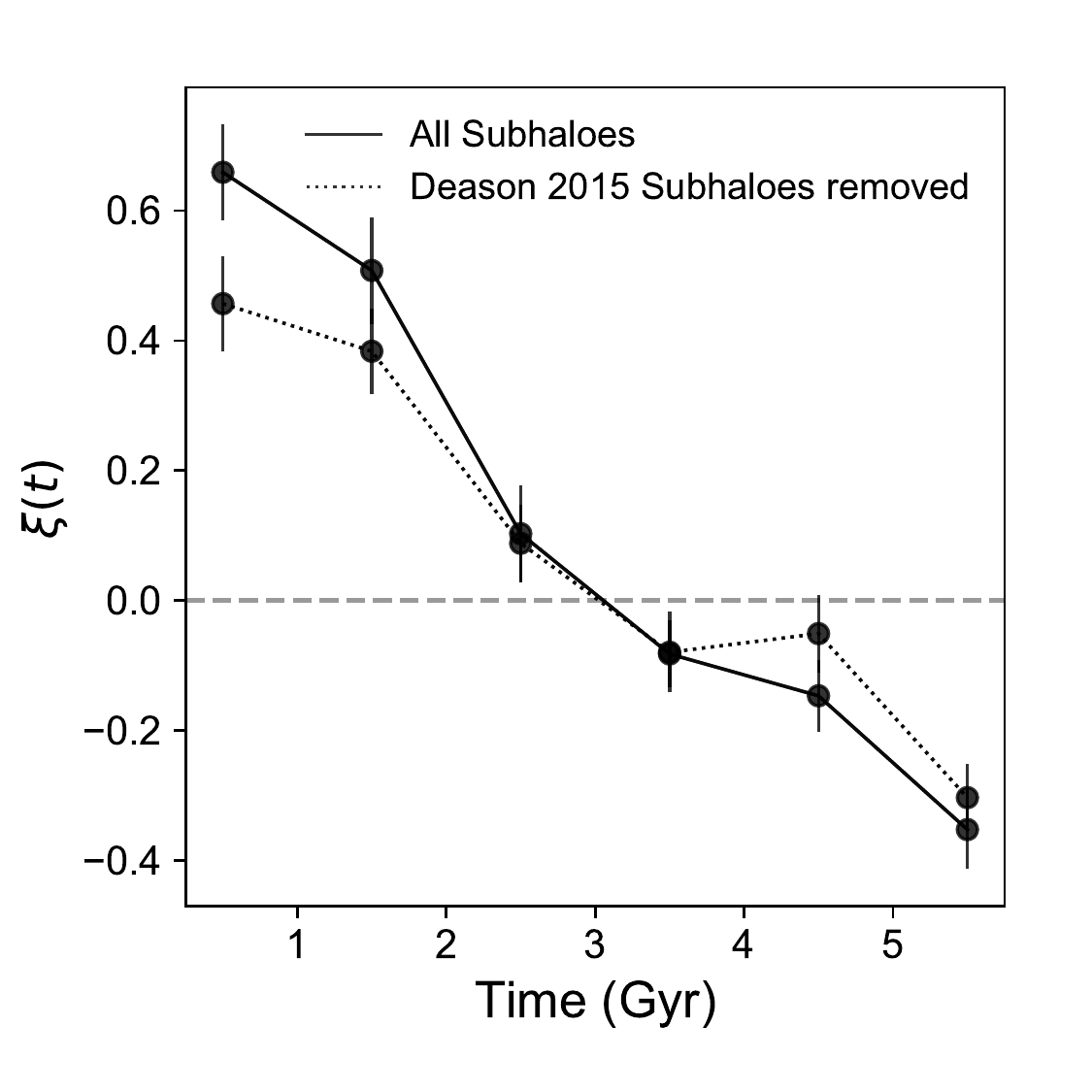}
    \caption{Destruction of subhaloes by massive progenitors: Cross-correlation function between the time when the massive progenitor first crosses a galacto-centric distance of 80 kpc from the MW-mass host and the destruction time of all subhaloes. Also shown is the same cross-correlation function omitting the subhaloes tagged to the massive progenitor according to the Deason et al. criterion.}
    \label{fig:fig16}
	\end{center}
\end{figure}

We also attempt to quantify the destruction of subhaloes using a cross-correlation function. As in Section \ref{sec:3destruct}, we associate the time of destruction of a subhalo as the time when it merges with the main progenitor branch of the MW-mass halo. To calculate the cross-correlation function of the destruction time of subhaloes, we use a similar estimator as above and a random sample generated from the distribution of the disruption times of all destroyed subhaloes. We find that referencing to a  time when the massive progenitor first crosses a galacto-centric distance of 80 kpc maximises the cross-correlation signal for the destruction of subhaloes. At these distances, the massive progenitor begins to influence the orbits of subhaloes inside the virial radius of the MW-mass halo. We adopt this fiducial crossing-time for calculating the cross-correlation function of destruction of subhaloes in Fig. \ref{fig:fig16}. Moreover, there is a slow decrease in the cross-correlation function with time. This reflects our intuition from Section \ref{sec:3destruct} that the destruction of subhalos is not immediate but happens gradually with time. Finally, even removing the subhaloes tagged by the Deason et al. criterion, there is still a large cross-correlation signal, suggesting that the clustering in the cross-correlation function is due not only to the destruction of a number of subhaloes belonging to the massive progenitor, but also due to the destruction of pre-existing earlier-accreted subhaloes whose orbits have been disturbed by the massive progenitor.

\begin{figure*}
	\begin{center}
    \includegraphics[width=0.8\textwidth]{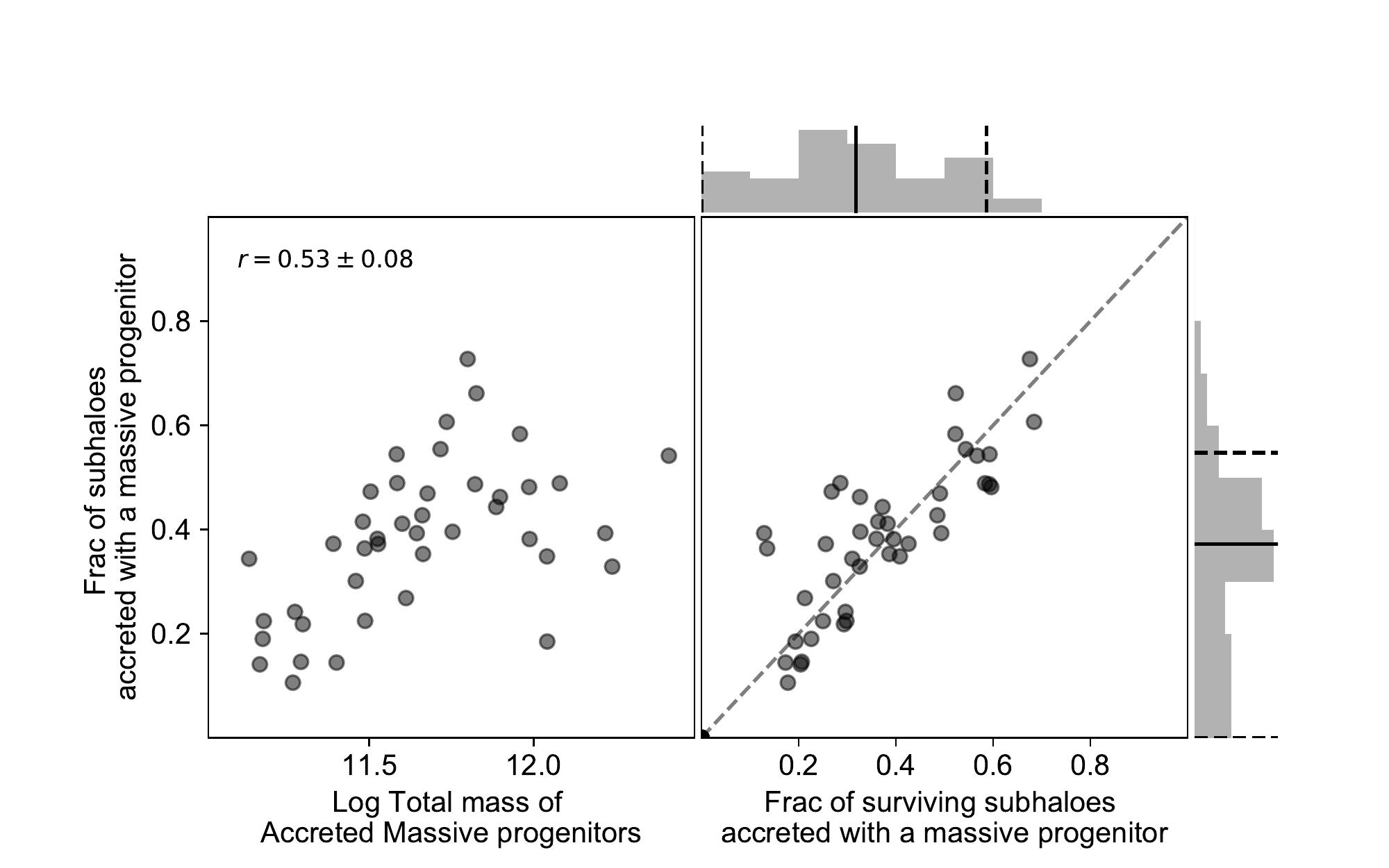}
    \caption{{\it Fraction of subhaloes accreted along with a massive progenitor}. Left: The fraction of subhaloes (surviving+destroyed) accreted along with a massive progenitor as a function of the total mass of the massive progenitors. Right: The fraction of subhaloes (surviving+destroyed) accreted along with a massive progenitor as a function of the fraction of surviving subhaloes. The fraction of subhaloes accreted along a with massive progenitors increases with the mass of the massive progenitors.}
    \label{fig:fig17}
	\end{center}
\end{figure*}

We can now attempt to quantify the fraction of subhaloes accreted along with a massive progenitor in MW-mass haloes. Following Section \ref{sec:3} and now reinforced by Fig. \ref{fig:fig15}, we consider that there is a temporal association between a subhalo and a massive progenitor if they were accreted within a Gyr of each other. In Fig. \ref{fig:fig17}, we find that the fraction of subhaloes accreted along with a massive progenitor increases with the total mass of the massive progenitors accreted by the MW-mass halo, and ranges between 10 and 70\%. This correlation is due to two reasons. Firstly, more massive progenitors tend to bring in a larger number of subhaloes. Secondly, the more the number of massive progenitors, the higher will be the associated fraction of subhaloes. Furthermore, the fraction of surviving and accreted subhaloes accreted along with a massive progenitor is about the same as the fraction of surviving subhaloes with some scatter (see the right panel of Fig.\ \ref{fig:fig17}).

\begin{figure}
	\begin{center}
    \includegraphics[width=0.8 \columnwidth]{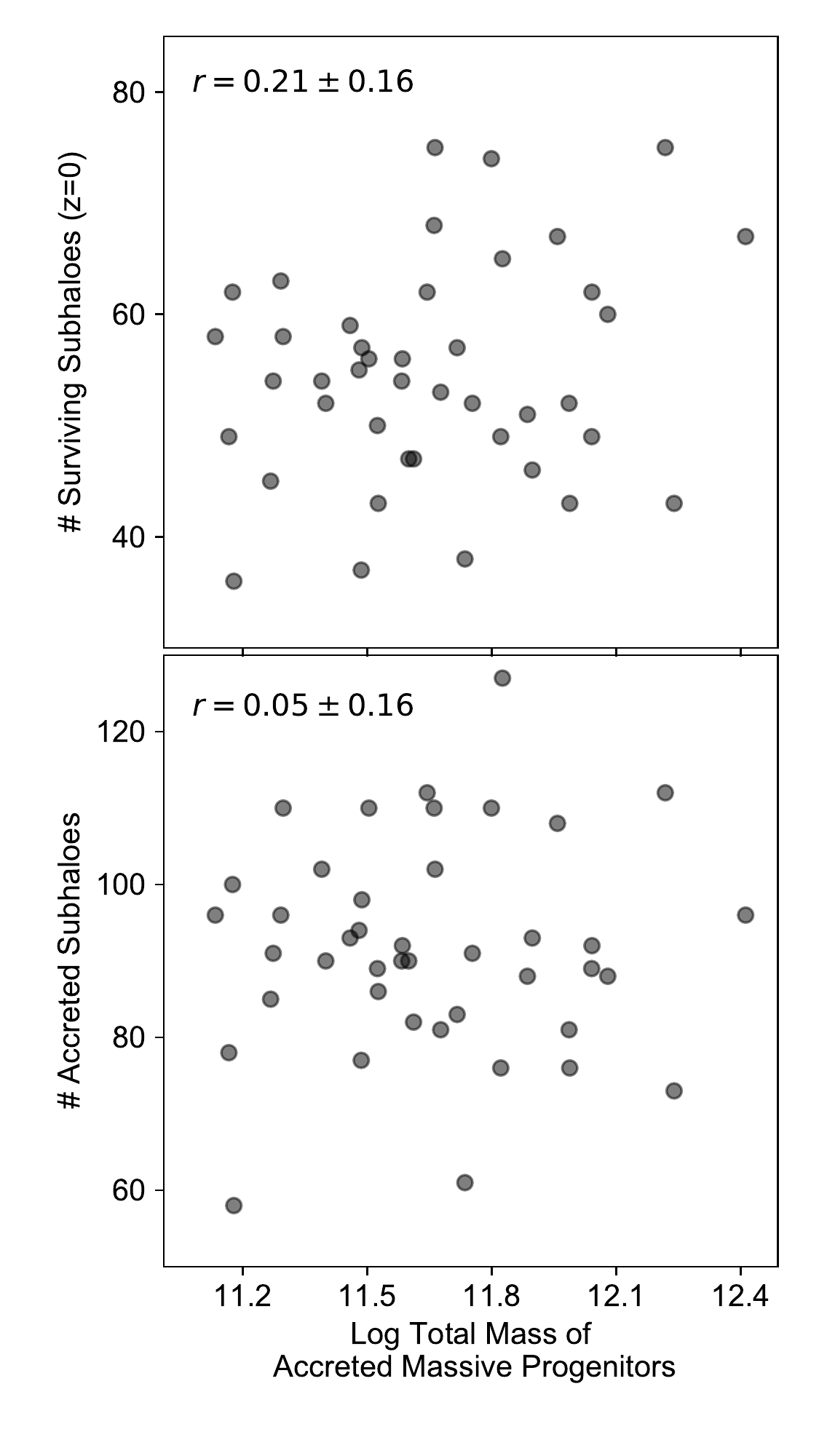}
	\caption{The number of surviving subhaloes (top) and the total number of accreted subhaloes (bottom) as a function of the total mass of the accreted massive progenitors. The Pearson correlation coefficient (r) is calculated, while its uncertainty is estimated using bootstrapping. No correlation is found between the total mass of the massive accretions and the number of surviving subhaloes or the total number of accreted subhaloes.}
	\label{fig:fig18}
	\end{center}
\end{figure}

While the fraction of subhaloes accreted along with a massive progenitor increases with the mass of the latter, the final number of surviving subhaloes in a MW-mass host shows no dependence on the mass of the massive progenitor. We demonstrate this in Fig. \ref{fig:fig18} where we find no correlation between the total mass of the massive accretions the total number of accreted subhaloes (surviving and destroyed). The correlation between the total mass of the massive accretions and the number of surviving subhaloes is also insignificant. This result, while surprising, reinforces two fundamental ideas: a) that the number of infalling subhaloes onto a MW-mass host is primarily a function of its own total mass and b) the number of surviving subhaloes is shaped by the time of the infall of the subhaloes onto the MW-mass host, with subhaloes accreted earlier tending to be destroyed faster \citep[e.g.][]{Ishiyama2008}. Massive accretions, therefore, do not significantly increase the number of infalling subhaloes, but instead serve to cluster the infall and destruction of subhaloes in time.  

\begin{figure}
	\begin{center}
    \includegraphics[width=0.8 \columnwidth]{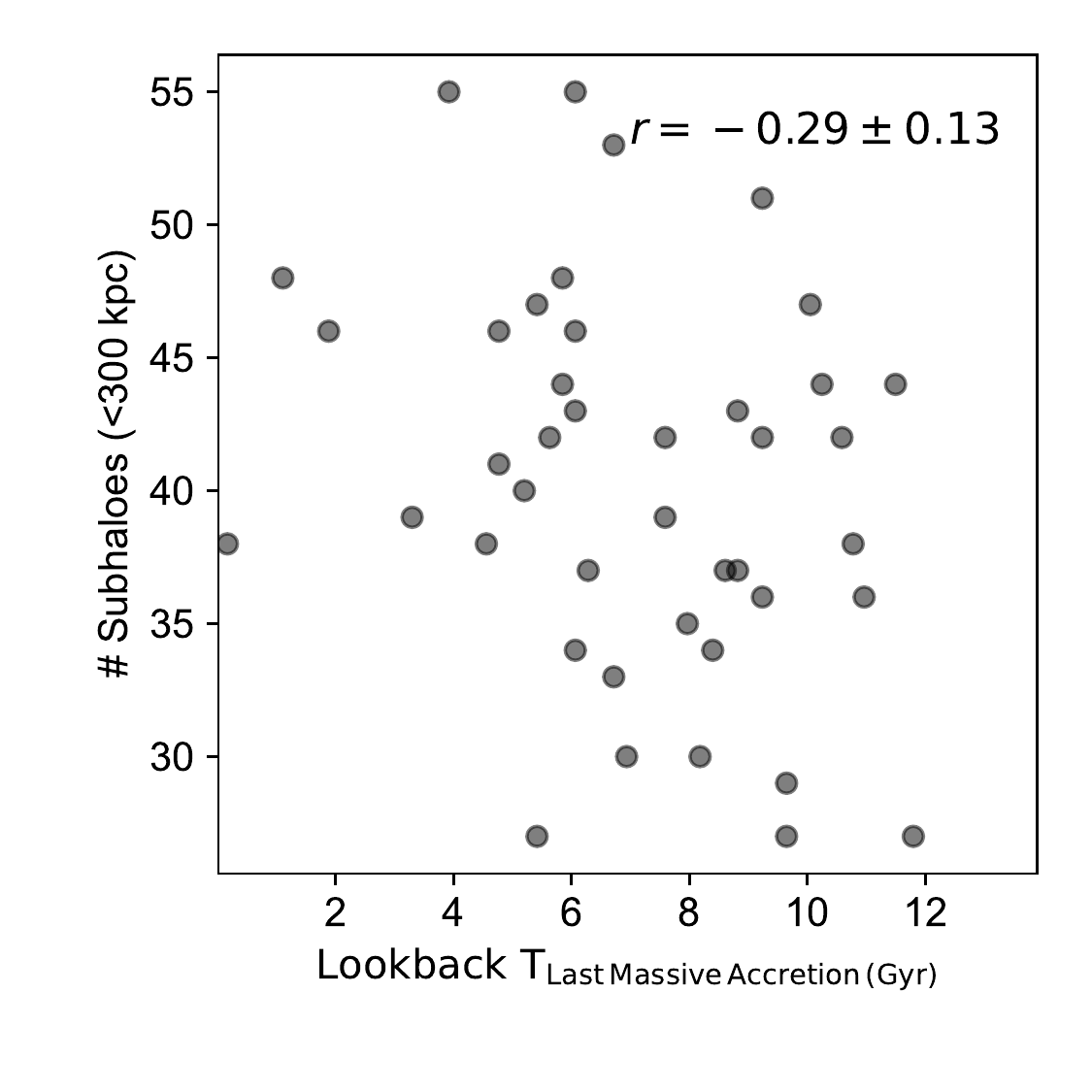}
	\caption{The number of surviving subhaloes of MW-mass (within <300 kpc) anti-correlates with the time of accretion of the last massive progenitor (p-values=0.01)}
	\label{fig:fig19}
	\end{center}
\end{figure}

This ability of massive progenitors to cluster the infall of subhaloes leads to a subtle temporal correlations with the number of surviving subhaloes. In Fig. \ref{fig:fig12}, we demonstrated that the infall of a massive accretion tends to increase and then decrease the number of subhaloes with cosmic time. MW-mass haloes which suffered a recent accretion will have an elevated number of subhaloes. Furthermore, the time of accretion of the most recent progenitor correlates with the median time of infall of all subhaloes, and thus the number of surviving subhaloes. Therefore, we should expect a subtle anti-correlation between the number of surviving subhaloes of a MW-mass halo (within a certain galacto-centric radius) and the time of accretion of the most recent massive progenitor, which we present in Fig. \ref{fig:fig19}. Since massive accretions are an important part of halo growth, the infall time of the most massive accretions correlates also with halo formation time and concentration. Accordingly, Fig.\ \ref{fig:fig19} echoes the well-known anti-correlation between the number of surviving subhaloes and the halo concentration or formation time reported elsewhere \citep[e.g.,][]{Zentner2005,Zhu2006,Ishiyama2008}.

\subsection{Orbits}
\begin{figure}
	\begin{center}
    \includegraphics[width=0.8\columnwidth]{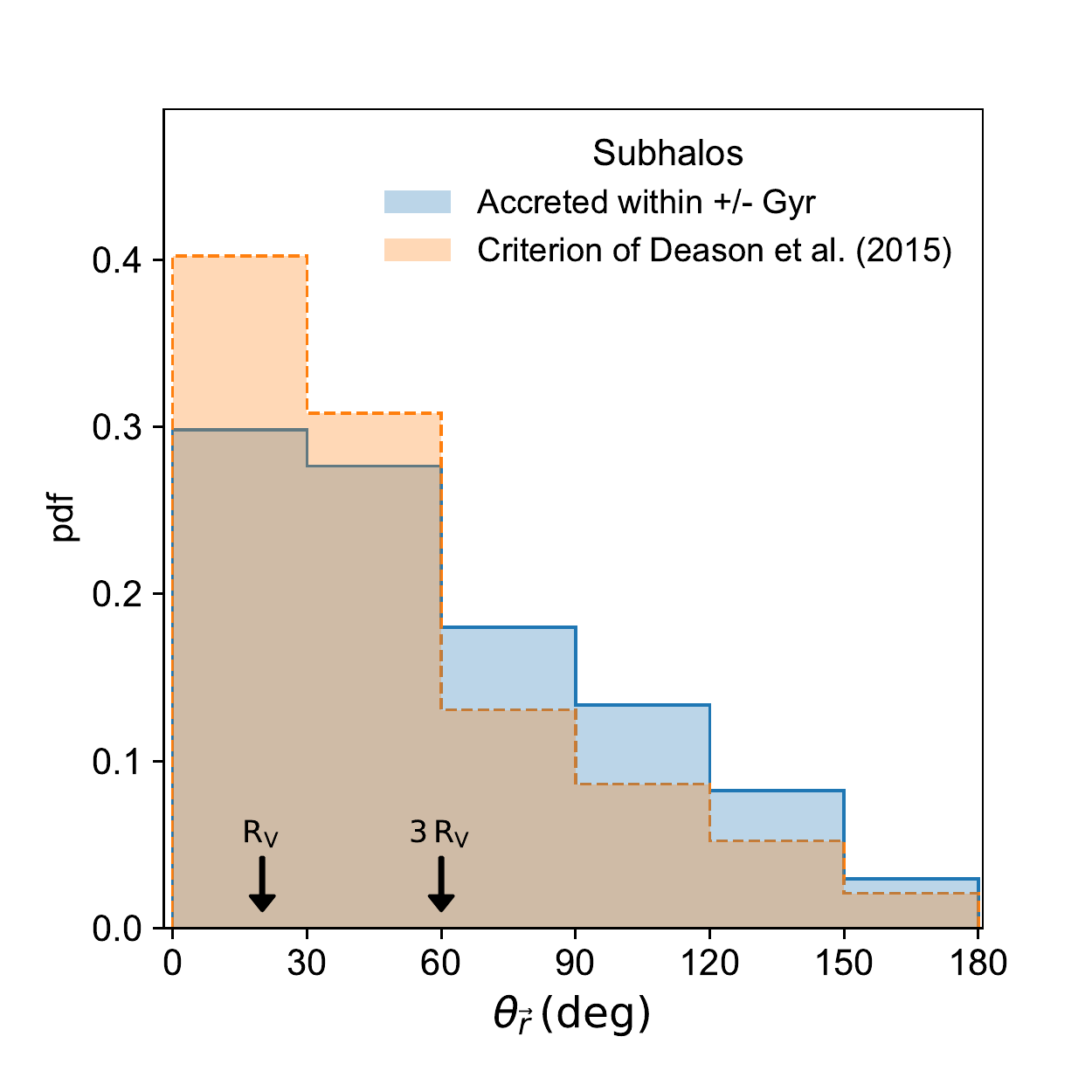}
    \caption{The pdf of the angles between the positional vector of the most recent massive accreted progenitor and its subhaloes in the `pre-accretion phase' (2 Gyr before accretion). Subhaloes are chosen in two ways: a) Subhaloes accreted within +/- 1 Gyr of the massive accreted progenitor, b) according to the criterion of Deason et al. 2015. Also marked are the angular size of the virial radius at these distances.}
    \label{fig:fig21}
	\end{center}
\end{figure}

We have already seen that the massive progenitor attracts subhaloes towards itself before infall into the MW-mass host. Hence the subhaloes accreted along with a massive progenitor appear to be attracted towards the MW-mass host from a large solid angle in the sky. To demonstrate this, we plot the probability distribution function (pdf) of the angles on the sky between the positional vector of the most recent massive progenitor and the subhaloes accreted along with it 2 Gyr before its accretion into the MW-mass halo, a time when the massive progenitor is sufficiently isolated and is not subject to strong tidal stripping or harassment. Furthermore, we only select MW-mass hosts which did not suffer another massive accretion within +/- 2 Gyr of its most recent massive accretion. We select subhaloes in two ways: a temporal association of 1 Gyr or according to the criterion of Deason et al. From Fig. \ref{fig:fig21}, we find that subhaloes appear to be accreted along with the massive progenitor from a range of relative angles in the sky from the massive progenitor, with the median angle being 50$^\circ$. Moreover 10 \% of the accreted subhaloes selected according to the temporal criterion of 1 Gyr have relative angles greater than 120$^\circ$. The accretion of massive progenitors concentrates both spatially and temporally the infall of subhaloes.

\begin{figure}
	\begin{center}
    \includegraphics[width=0.8 \columnwidth]{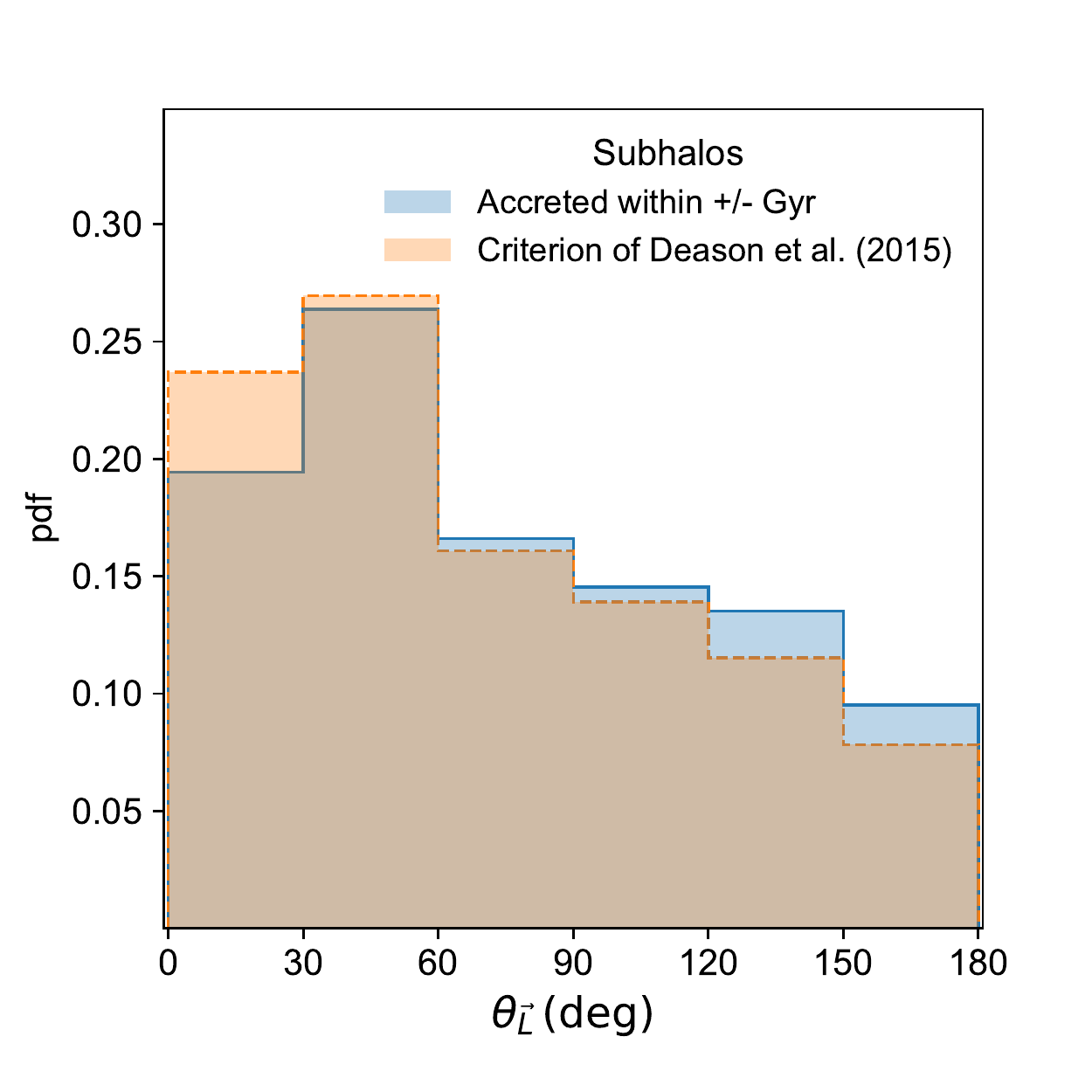}
	\caption{The angle between the orbital poles of the most massive accreted progenitor and the subhaloes accreted along with it $\sim$2.5 Gyr after accretion (post-accretion phase). Subhaloes were chosen in two ways: a) Subhaloes accreted within +/- 1 Gyr of the massive accreted progenitor, b) according to the criterion of Deason et al. 2015. }
    \label{fig:fig22}
	\end{center}
\end{figure}

The large size of the accreted group leads to a wide range in impact parameters and hence, in the `post-accretion' phase, the subhaloes accreted along with a massive progenitor may have a considerable diversity in their orbits. In order to understand whether the orbits of the surviving satellites can reveal the orbit of the massive progenitor, we plot the angle between the orbital poles of the last massive accretion and the subhaloes accreted along with it 2.5 Gyr after the time of accretion of the massive progenitor; we choose a time of 2.5 Gyr, as it is the typical median time for a massive progenitor to be destroyed after entry into the halo (see Appendix \ref{Appendix:1}). Again, we select only MW-mass hosts which did not suffer another massive accretion within +/- 2 Gyr of the last massive accretion. In Fig. \ref{fig:fig22}, we find that there is a large scatter in the orbital poles of the subhaloes with respect to that of the massive progenitor, with angles ranging from 0 to as far as 180 degrees. The subhaloes accreted along with the massive progenitor are not restricted to a plane; a small fraction are found rotating in the opposite direction to that of the massive accretion.

\section{Confrontation with Observations}
\label{sec:data}
In the previous section, we demonstrated how massive accretions serve to concentrate the accretion of subhaloes onto a MW-mass host causing a clustering in their time of infall. These subhaloes accreted along with a massive progenitor are predicted to have a range of orbits, making it difficult to isolate them solely based on their phase-space information. Thus, it remains a challenging task to identify the dwarf satellites accreted along with the MW and M31's massive accretions. However, if indeed the massive accretions did contribute a number of satellites to the MW and M31, then we should see a clear signature in the clustering of the infall time of the dwarf satellites.

Without direct access to the infall time of satellites, we are left with proxies like the quenching time of satellites. It is generally believed that dwarf galaxies ($10^{5} \mathrm{\msun} \lesssim \mathrm{M}_{*} \lesssim 10^{7} \mathrm{\msun}$) were quenched in the environment of their MW-mass hosts, preferably at the first pericenter passage \citep[e.g.][]{Slater2014}. However, in the last few years this understanding has been questioned \citep[e.g.][]{Rocha2012,Weisz2015,Fillingham2019}. For the MW and M31, where we have prior knowledge of the infall time of their massive progenitors, a clustering of the quenching time around the infall time of the MW and M31's massive progenitors would not only validate the mechanisms of quenching in these dwarf galaxies but will also be a critical test of the picture developed in this paper.

In this section, we study the temporal clustering of the lookback quenching times of the selected dwarf satellites of the MW and M31 around their known massive mergers. Following the literature \citep[e.g.][]{Weisz2015,Weisz2019b}, we adopt the time at which 90\% of the total stellar mass was formed ($\tau_{90}$) as a proxy for the lookback quenching time.

For this exercise, we consider only dwarf satellite galaxies of the MW and M31 in the magnitude range $ -13.6 < M_{\mathrm{V}} < -6 $  with measured star formation histories (e.g. \citealt{Weisz2014,Weisz2019b,Skillman2017,Bettinelli2018,Torrealba2016}).  Our choice of the magnitude range is motivated by two factors. First, we avoid bright dwarfs which may have continued their star formation long after they enter the halo of their host galaxy (e.g., Sagittarius, SMC, LMC, M33). Second, we also avoid ultra-faint dwarfs which may have be quenched early due to reionisation. Although it is conventional to consider  $M_{\mathrm{V}} < -7.7$ satellites as ultra-faint dwarfs \citep{Simon2019}, there is evidence that some low-mass dwarf galaxies ($-7.7 < M_{\mathrm{V}} < -6$) in M31 may have been quenched recently \citep{Weisz2019b}. For this reason, we also include dwarf galaxies brighter than $M_{\mathrm{V}} < -6$. Two galaxies in our sample (Fornax $M_{\mathrm{V}} \sim -13.5$ and Carina $M_{\mathrm{V}} \sim -9.5$) merit a special mention. Although both dwarf galaxies have been quenched fairly recently ($\sim 2.5-3$ Gyr ago), there are suggestions that these dwarf galaxies have been orbiting the MW for a much longer time \citep[e.g.][]{Pasetto2011, Rocha2012, Patel2020, Rusakov2020}. However, the results are dependent upon a number of model uncertainties and are much debated in the literature \citep{Pardy2019,Jahn2019}. We choose to retain Fornax and Carina in our sample. This leaves us with a sample of 13 and 28 satellites for the MW and M31 respectively (see Table \ref{table:table1}).

The lookback time of shutdown in star formation of these dwarf galaxies has been drawn from the literature, and has been estimated from data of varying quality. For 6 M31 dwarfs drawn from the ISLAandS project \citep{Skillman2017}, the shutdown time has been estimated from deep CMDs reaching down to the oldest main sequence turn-off (MSTO). The shutdown time for the vast majority of the M31 dwarfs have been estimated from resolved HST data reaching fainter than the horizontal branch \citep{Weisz2019b}. A sizeable fraction of the satellites of the MW have their time of their shutdown in star formation estimated from HST data reaching down to the MSTO \citep{Weisz2014}.  For a few of the nearer MW dwarfs, the star formation has been estimated using ground-based data \citep[e.g.][]{Bettinelli2018,Torrealba2016}. In Table \ref{table:table1}, we have included both the systematic and statistical uncertainties of $\tau_{90}$. The constraints on $\tau_{90}$ as well as the ancient star formation history of M31 dwarfs will dramatically improve with observations from the upcoming Cycle 27 HST Treasury survey as part of HST-GO-15902 (PI D. Weisz).

\begin{table}
\begin{center}
	\caption{Star Formation Shutdown Time}
	\begin{threeparttable}
		
		\begingroup
		\setlength{\tabcolsep}{10pt} 
		\renewcommand{\arraystretch}{1.3} 
\begin{tabular}{ l c c c l}
\hline
 Satellite Name & $M_{\mathrm{V}}$ & $D_{\mathrm{Host}}$ & $\tau_{90}$ \\
 & (mag) & (kpc) & Gyr \\
 \hline
Cas~III  & -12.6  & 141 &  $4.1_{-1.5}^{+2.5}$  \\
Cas~II  & -11.2  & 148  &   $7.2_{-3.4}^{+2.8}$  \\
And~XXIII  & -10.0 &  129 & $5.1_{-2.8}^{+1.5}$   \\
And~XXV  & -9.3  &  94 & $5.8_{-1.3}^{+2.6}$   \\
And~XXI  & -9.2  &  134 & $5.8_{-2.5}^{+0.9}$  \\
And~IX  & -9.0  &  186  & $5.1_{-2.0}^{+1.8}$   \\      
And~XIV  & -8.6  &  161  & $4.8_{-0.7}^{+5.2}$   \\
And~XXIX  & -8.5  &  188  & $5.2_{-1.2}^{+2.2}$   \\
And~XVII  & -8.2  &  70  & $10.5_{-5.0}^{+2.1}$  \\
And~XXIV  & -7.9  &  167  & $5.4_{-3.1}^{+4.4}$  \\
And~II  & -12.6  & 198  & $6.3_{-0.6}^{+0.5}$  \\
And~I  &  -12.0  & 70 & $7.4_{-0.7}^{+0.9}$  \\
And~III  &  -10.1 &  88 & $8.7_{-0.6}^{+1.5}$  \\
And~XV  &  -8.4  &  179 & $9.3_{-0.8}^{+3.3}$  \\
And~V  &  -9.1 & 115  &  $7.413_{-2.3}^{+1.7}$ \\
And~VI  &  -11.3 & 269  &  $5.37_{-0.5}^{+1.6}$ \\
And~VII  &  -12.6 & 220  &  $10_{-1}^{+1}$ \\
Lac~I  &  -11.5 & 265  &  $4.9_{-1.7}^{1.7}$ \\
Per~I  &  -10.2 & 353 & $4.0_{-1.6}^{2.6}$ \\
And~XVIII  &  -9.2 & 453 & $4.6_{-2.1}^{1.7}$ \\
And~X  &  -7.5  & 137 &  $6.5_{-2.1}^{4.8}$ \\
And~XII  &  -7.0 & 179 & $3.4_{-0.2}^{2.6}$ \\
And~XXII  &  -6.7 & 274 & $6.8_{-2.5}^{5.8}$ \\
And~XX  &  -6.7 & 130 & $6.9_{-2.1}^{4.6}$ \\
And~XIII  &  -6.5 & 132 & $6.5_{-0.1}^{3.5}$ \\
And~XI  &  -6.3 & 111 & $7.4_{-1.4}^{2.4}$ \\
And~XXVI  &  -6.1 & 103 & $9.1_{-6.0}^{2.9}$ \\
And~XXVIII  &  -8.8 & 368 & $7.6_{-0.3}^{1.7}$ \\

\hline
Fornax & -13.5 & 149 & $2.4_{-0.2}^{+0.3}$ \\
Carina & -9.1 & 107 &  $2.9_{-0.3}^{+0.3}$ \\
Canes~Venatici~I & -8.6 &  218 & $8.3_{-0.9}^{+0.5}$ \\
Draco & -8.8 & 76 &  $10.2_{-1.1}^{0.6}$ \\
Leo~I & -12.0 & 257 &  $1.7_{-0.0}^{0.1}$ \\
Leo~II & -9.8 & 236 &  $6.5_{-0.3}^{0.3}$ \\
Sculptor & -11.1 & 86 &  $10.7_{-1.8}^{0.5}$ \\
Ursa~Minor & -8.8 & 78 &  $9.1_{-1.6}^{0.6}$\\
Sextans & -9.3 & 89 &  $11_{-0.5}^{0.5}$ \\
Crater~2 & -8 & 116 &  $11_{-0.5}^{0.5}$ \\ 
Hercules & -6.6 & 126 &  $10.2_{-6.2}^{0.0}$ \\
Bootes~I & -6.3 & 64 &  $11_{-0.5}^{0.5}$ \\
\hline
\end{tabular}
\endgroup
\begin{tablenotes}
\item Distance of M31 satellites is taken from \cite{Conn2012} and \cite{Weisz2019a}. $\tau_{90}$ is taken \cite{Weisz2014}, \cite{Weisz2019b}, \cite{Brown2014}, \cite{Bettinelli2018}, \cite{Torrealba2016}. $\mathrm{M_{V}}$ taken from \cite{McConnachie2012}.
\end{tablenotes}
\end{threeparttable}
\label{table:table1}
\end{center}
\end{table}

\subsection{Temporal Clustering of Star Formation Shutdown}
\begin{figure}
\begin{center}
	\includegraphics[width=0.8\columnwidth]{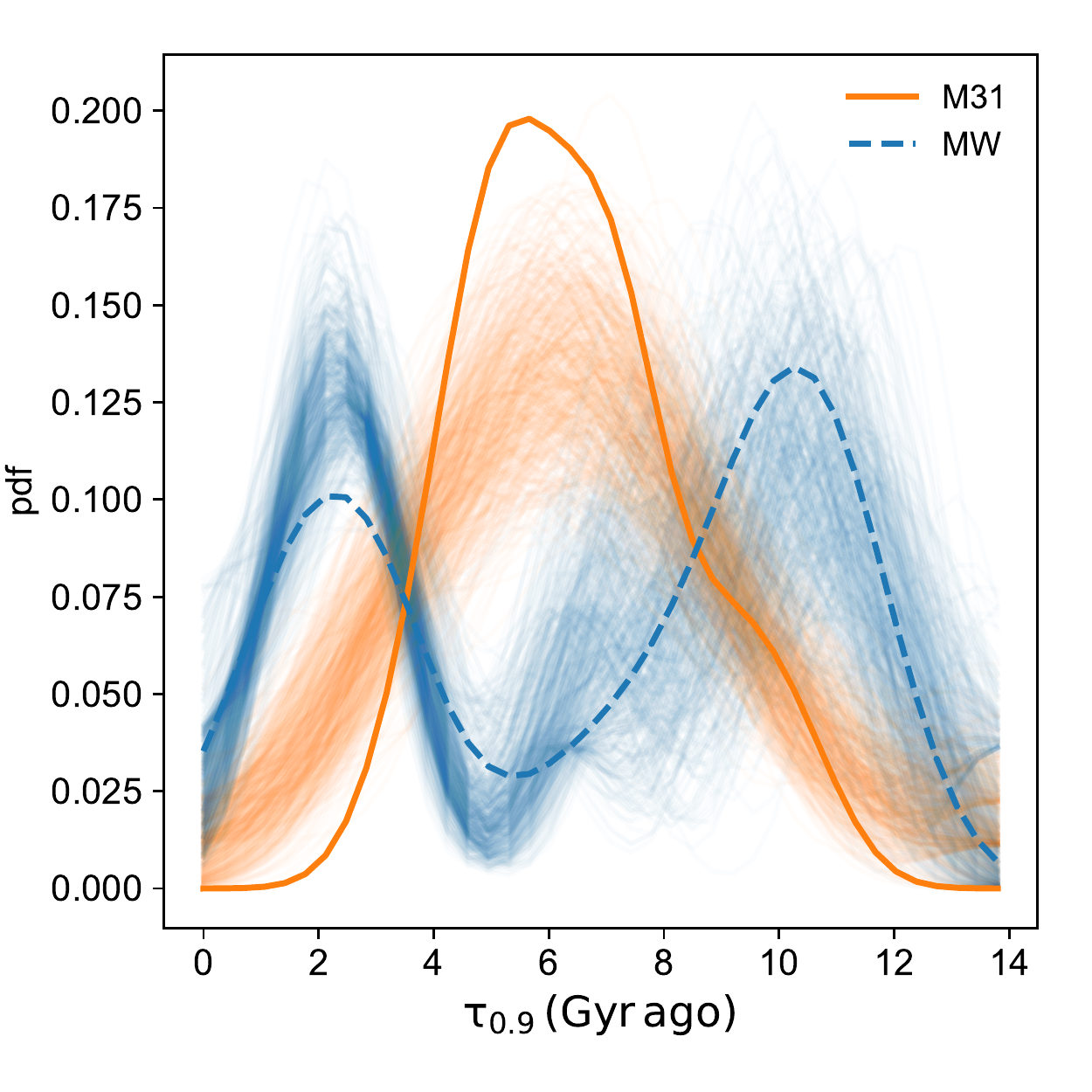}
    \caption{The probability density functions (pdf) of the lookback time of the shutdown of star formation ($\tau_{0.9}$) of the MW and M31, estimated through a kernel density estimator. The single thick solid lines show the pdf using the given data without their uncertainties. The thin solid lines show a Monte Carlo distribution of the pdfs taking into consideration the uncertainties in the time of accretion. The quenching times of the satellites of the MW and M31 are strongly clustered, suggesting that a substantial number of them were accreted in groups. The inclusion of the uncertainties for the MW favours the recent peak around 2 Gyr ago over the second peak around 10 Gyr ago.}
 \label{fig:fig25}
 \end{center}
\end{figure}

The lookback time of shutdown in star formation in the satellites of the MW and M31 is strongly clustered. We demonstrate this by employing a kernel density estimator (KDE) to estimate the mean lookback quenching time of these clusters. In the bottom panel of Fig. \ref{fig:fig25}, we estimate the the probability density function (pdf) of the lookback quenching time. To do this, we select the `bandwidth' (smoothing parameter) for each KDE, such as to to optimise the bias-variance tradeoff (1.29 and 0.98 for the MW and M31 respectively). The pdf of the MW is double humped with prominent peaks at $\sim2.5$ Gyr and $\sim11$ Gyr ago. The pdf of M31 has a single prominent peak $\sim6$ Gyr ago. These prominent peaks continue to be significant and stable, even after taking into account the uncertainties in the lookback quenching time of the satellites. The inclusion of the uncertainties for the MW favours the recent peak around 2 Gyr ago over the second peak around 10 Gyr ago. The full width half maximum of the pdfs around the peaks account for 77 and 62 percent of the satellites in the MW and M31 respectively. The higher values for the MW are probably caused by larger bandwidth used for the KDE but also due to the presence of a second peak. Under the assumption that the lookback quenching time strongly correlates with the infall time (entry) of the satellites into the MW-mass halo, this suggests that the satellites of the Local Group were not accreted uniformly over time, but a substantial number of the them were accreted in groups.

\subsection{Strong Correlation between large accretions and shutdown in star formation in dwarf satellites}
\begin{figure}
\begin{center}
	\includegraphics[width=0.8\columnwidth]{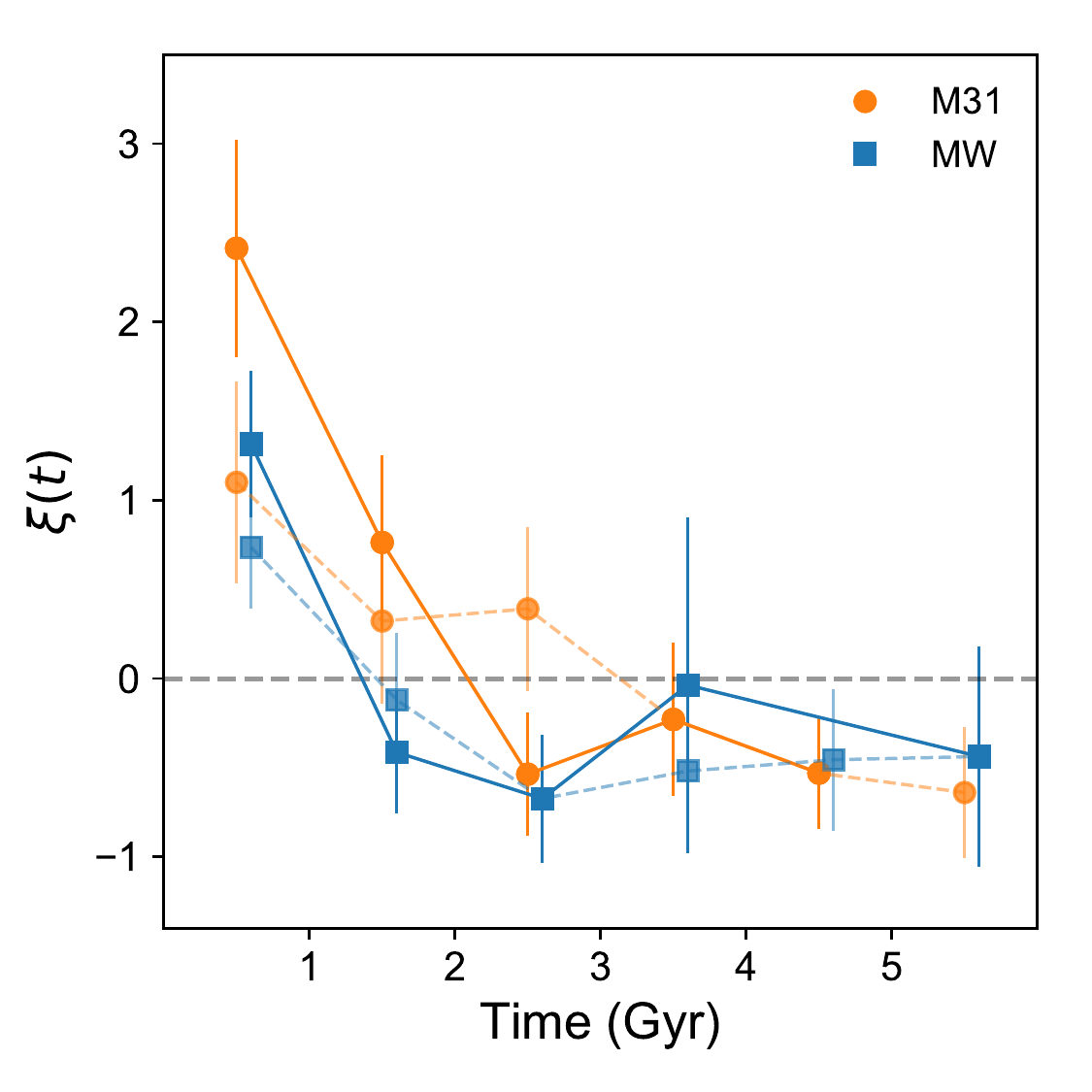}
    \caption{The cross-correlation function between the accretion time of large accreted progenitors of the MW/M31 and the quenching lookback time of the satellites of the MW and M31. We consider M31's massive accretion at 5.5 Gyr ago and MW's accretion of the LMC and Gaia-Enceladus 2.5 and 10.5 Gyr ago respectively. As in Fig. \ref{fig:fig1}, the solid lines represent a cross-correlation function which does not account for the uncertainties in the shutdown time, while the dashed lines does.} 
 \label{fig:fig26}
 \end{center}
\end{figure}

Given our present knowledge of the large accretions suffered by the MW and M31, we calculate the cross-correlation function between the estimated infall time of these large accretions and the lookback quenching time of the satellites of the MW and M31. There is firm evidence that the MW has suffered at least two large accretions, the LMC and Gaia-Enceladus \citep{Helmi2018}. It is also believed to have suffered a number of smaller but significant accretions (e.g. Sagittarius). On the other hand, M31 is believed to have merged with a large galaxy, whose tidal debris now makes up a significant fraction of M31's large stellar halo \citep{DSouza2018b, Hammer2018}. There is considerable debate about the accretion time of M33 into the halo of the M31 \citep[e.g.][]{McConnachie2009,Patel2017,Semczuk2018,TepperGarcia2020}. For the purpose of this paper, we consider only the two large accretions of the MW and a single large accretion for M31. Furthermore, it is to be noted that our constraints on the time of accretion of these massive progenitors are subject to large model uncertainties. The expected crossing-time of the LMC into the virial halo of the MW depends on their assumed masses, and is about $\sim1.5-3$ Gyr \citep{Kallivayalil2013}. The accretion of Gaia-Enceladus is estimated to be about $\sim10-11$ Gyr \citep{Helmi2018}. On the other hand, M31's large accreted progenitor is expected to have entered the virial halo about $\sim5-6$ Gyr ago (see Appendix \ref{Appendix:1}).  We adopt the following accretion times for the MW: 2.5 and 10.5 Gyr for the LMC and Gaia-Enceladus respectively. For M31, we adopt an accretion time of 5.5 Gyr for its massive progenitor. In Fig. \ref{fig:fig26}, we calculate the cross-correlation function between these fiducial accretion times and the lookback quenching time of the satellites of the MW and M31. We find that the quenching time of the satellites of the Local Group are strongly correlated with the the accretion time of the large massive progenitors. This suggests that a significant fraction of the satellites of both the MW and M31 were brought in by their large accretions and lends weight to the idea that the quenching time is a good proxy for their infall times for the dwarf satellites considered in this section.

The cross-correlation function of the quenching time of the satellites of M31 is subject to the uncertainty of the accretion time of M33. If the accretion time of M33 is between $\sim4$ and $\sim9$ Gyr ago and overlaps with the quenching time of the bulk of M31's satellites, the results of cross-correlation function will not significantly change the results. On the other hand, an accretion time of M33 of less than 2 Gyr ago would reduce the central amplitude of the cross-correlation function by 50\%. Better constraints on the infall time of M33 are needed to reduce the uncertainties of the cross-correlation function of quenching time of satellites in M31.

\section{Conclusions and Discussion}
\label{sec:conclusions}
In this work, we demonstrate that infall of a massive progenitor onto a MW-mass host is accompanied by the accretion and destruction of a number of subhaloes capable of hosting dwarf satellite galaxies. Massive accreted progenitors do not increase the number of infalling subhaloes onto a MW-mass host but instead serve to cluster their time of infall and destruction. Apart from contributing their own established subhaloes, massive accretions also concentrate surrounding subhaloes onto the MW-mass host, leading to a temporary elevation in the number of subhaloes. The concentration of the cumulative radial profile of the subhaloes changes with respect to the position of the massive progenitor around the MW-mass host. Surviving associated subhaloes with a massive progenitor have a large diversity in their orbits and are not restricted to a thin plane. Finally, we show that the clustering in quenching time of dwarf spheroidal galaxies in the MW and M31 around their known massive accretions is consistent with the expected clustering of infall times, suggesting that a significant fraction of their dwarf satellites fell in along with the massive progenitors.

\subsection{Limitations}
This work has a number of limitations connected with its use of subhaloes of DM-only simulations to infer the properties of dwarf satellite galaxies. First, there is a degree of uncertainty about which subhaloes are capable of hosting dwarf satellite galaxies. Second, it does not take into account the selective destruction of subhaloes on radial orbits due to the disk of host galaxy. We attempt to address these limitations and demonstrate that they do not affect the main conclusions of this paper.

First, we find that the temporal clustering properties of the infall of subhaloes likely to host satellites around massive accretions is not very different from the clustering properties of a wider set of subhaloes around massive accretions. To illustrate this, we implement a simple semi-analytical scheme to populate subhaloes with `satellites', which are outlined in Appendix \ref{Appendix:2}. In Fig.\ \ref{fig:fig30}, we plot the cross-correlation function of the infall time of subhaloes chosen to host `satellites' with the infall time of massive accretions. We find that that there is no statistical difference between this cross-correlation function and the cross-correlation function considering all subhaloes. Therefore, we conclude that infall properties of all with $\mathrm{M_{\rm peak}^{\prime} > 10^9 \msun}$ subhaloes are similar to those of the subhaloes most likely to host dwarf satellite galaxies.

\begin{figure}
	\begin{center}
	\includegraphics[width=0.8\columnwidth]{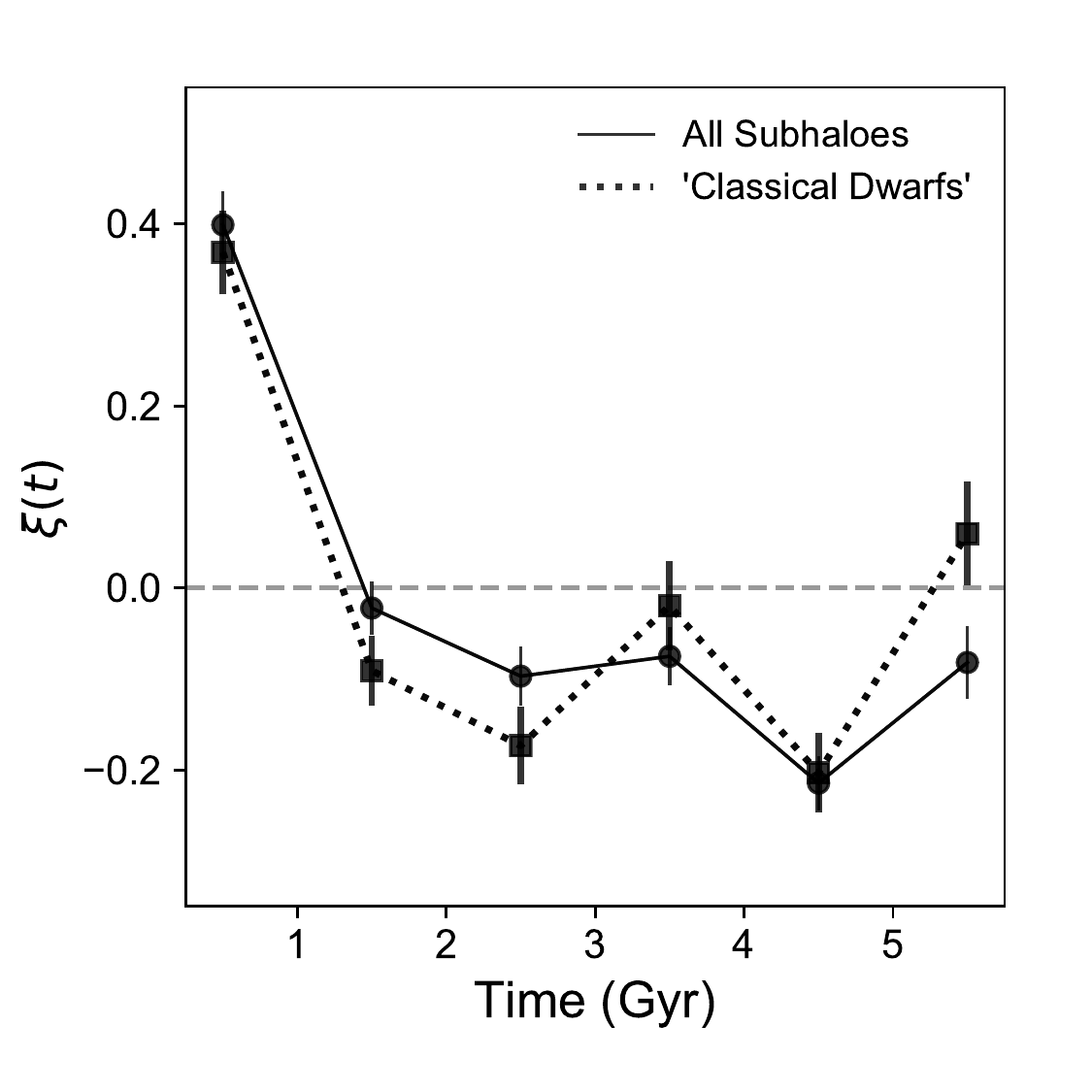}
    \caption{Cross-correlation function of the infall time of massive progenitors with the infall time of subhaloes possibly hosting `Classical Dwarfs'.}
    \label{fig:fig30}
	\end{center}
\end{figure}

Subhaloes are destroyed by galactic discs: \cite{Nadler2018} and \cite{Kelley2019} have demonstrated that subhaloes accreted more than 6.5 Gyr ago and that have suffered a number of close pericentric passages (<50 kpc) are preferentially destroyed by the disc of a MW-mass galaxy. This is an important issue, and we emphasise that very high resolution hydrodynamical models will be necessary to study the full life-cycle of satellite accretion, orbital evolution and their eventual destruction. Yet, a sizeable set of MW-mass haloes modelled using hydrodynamics with extremely high resolution is not readily available. Instead, we have chosen to focus on the infall phases of subhalo evolution {\it before} the disk has its most important impacts. Owing to this narrow focus, we do not expect any important limitations in our adoption of dark matter-only simulations for this work. This view is supported by comparison of normalised cumulative radial profiles of subhaloes in dark matter-only simulations and dwarf satellites ($\mathrm{M_{v}}>-9$) in hydrodynamical simulations --- which is sensitive not only to infall but also to satellite processing --- by \citet{Carlsten2020}, finding that there are no appreciable differences between their normalised cumulative radial profiles. In our work we have focused in an even more conservative set of measures, taken at infall or shortly afterwards, before differences in the potential between dark matter only and hydrodynamical simulations have had a chance to accumulate in their effects. We stress that even if we restricted our analysis to infall events in the last 6.5\,Gyr --- those events where \cite{Nadler2018} and \cite{Kelley2019} expect the smallest effects of a galactic disc --- our results are unchanged. We conclude that the conclusions of this work regarding the infall of subhaloes around massive accretions can be generalised safely to the infall of dwarf satellite galaxies.

Finally, we note that a number of subhaloes that were `previously bound' to the massive progenitor have completed a number of pericentric passages around it. It is probable that a number of these `previously bound' subhaloes will have their star-formation properties affected by the massive progenitor, a few Gyrs before being accreted onto the central MW-mass host. This will potentially affect the clustering signal in the star-formation shutdown times of dwarf galaxies. However, we note that a large fraction of these `previously bound' subhaloes are eventually destroyed, leaving a significant clustering in infall times of subhaloes hosting dwarf galaxies whose star-formation properties have not been modified by the massive progenitor.

\subsection{Applicability to the Nearby MW-mass Galaxies}
We now turn our attention to discuss how our results could be applicable to nearby MW-mass galaxies. We restrict our attention to the MW, M31, M81 and Cen A. The last two galaxies are motivated by the fact that we some prior knowledge about their accretion history. M81 has a small \citep{Harmsen2017,Smercina2020} and old stellar halo \citep{Durrell2010}, but is currently undergoing a massive accretion of two large galaxies (M82 and NGC3077). Though the orbital properties of M82 and NGC 3077 are unclear, it is believed that these progenitors have recently been accreted \citep[e.g.][]{Yun1999,Oehm2017}. In some respects, it parallels the ongoing accretion of the MW. On the other hand, we expect that Cen A should closely parallel M31; it too has a large stellar halo \citep{Rejkuba2014,DSouza2018a} which contains 10-20\% of 2-4 Gyr old stars, suggesting it had a recent accretion which was completely destroyed. Furthermore, the presence of many Sagittarius-like stellar streams found around M31 \citep{Ibata2014,McConnachie2018} and Cen A \citep{Crnojevic2016} further hint to their recent massive mergers.

First, in our analysis of Section \ref{sec:data}, we neglected the infall of M33 into M31 because of the uncertainty in its time of accretion. \cite{Patel2017} have suggested that M33 is on its first passage with its time of infall less than 2 Gyr; a suggestion which has been strengthened by improved measurements of the transverse velocity of M31 from Gaia data \citep{vanderMarel2019}. However, models which account for dynamical mass loss suggest a much earlier accretion history \citep[> 6.5 Gyr ago;][]{TepperGarcia2020}. The presence of the tidal features in the outskirts of M33 suggest that it already interacted with a much larger galaxy (presumably M31) in the past \citep[e.g.][]{Bekki2008, McConnachie2009, Semczuk2018}. The biggest uncertainty in the dynamical models is the transverse velocity of M31 and the masses of both M31 and M33. The quenching history of the classical dwarf satellites of M31 may give us an insight into this problem. In particular, dwarf satellite galaxies associated with the infall of M33 would bear the quenching signature of its time of accretion. From Fig. \ref{fig:fig25}, there is no sign of a significant peak in the quenching times of M31 satellites. Consequently, if M33 recently accreted in the last 2 Gyr and is on its first infall, it had very few satellites; alternatively, we would suggest that it is possible that M33's accretion time is earlier and it is not on its first infall.

Second, we showed in Fig. \ref{fig:fig14} that the concentration of the cumulative radial distribution of satellites is influenced by the position of the massive accreted progenitor on its orbit. While other effects also contribute to the radial distribution of satellites, massive accretions coming towards their first pericenter temporarily enhance the number of satellites within the inner parts of the MW host's halo. It has been noted that the cumulative radial distribution of the MW is more concentrated than that of M31 \citep[e.g.][]{Samuel2020a,Carlsten2020}. This may be easily interpreted as a reflection of the LMC's recent first pericentric passage. Similarly, the less concentrated projected radial distribution of M81 satellites \citep{Carlsten2020} can be explained if M82 is not close to its pericenter.

Third, the number of surviving satellite galaxies found within the virial radius of a MW-mass host is elevated immediately after the accretion of a massive progenitor. Although the dominant driver of the number of surviving satellites is the halo mass of the galaxy, we can still make some interesting observations. M81, whose stellar mass is not much greater than M31, has a large number of satellites. A possible explanation would be that the recent accretion of M82 and NGC3077 has elevated its satellite numbers. However, it is certain that the recent massive accretions of the M81 as well as the MW will eventually precipitate the destruction of a number of their satellites, thus lowering their total number of surviving dwarf satellites.

\section*{Acknowledgements}
We are grateful to Shea Garrison-Kimmel for providing us access to the catalogues and merger trees of the ELVIS simulations. E.F.B. is grateful for support from the National Science Foundation through grant NSF-AST 2007065 and NASA grant NNG16PJ28C through subcontract from the University of Washington as part of the \textit{WFIRST} Infrared Nearby Galaxies Survey.

\section*{Data Availability}
The data underlying this article are available in the article and is taken from the literature. The catalogues and the merger trees of the simulations analysed in this article were provided by Shea Garrison-Kimmel. These will be shared on request to the corresponding author with permission of Shea Garrison-Kimmel.



\bibliographystyle{mnras}
\bibliography{satinfall}



\appendix

\section{Estimating the time of accretion of M31's massive accreted progenitor}
\label{Appendix:1}
\begin{figure}
	\begin{center}
	\includegraphics[width=0.8 \columnwidth]{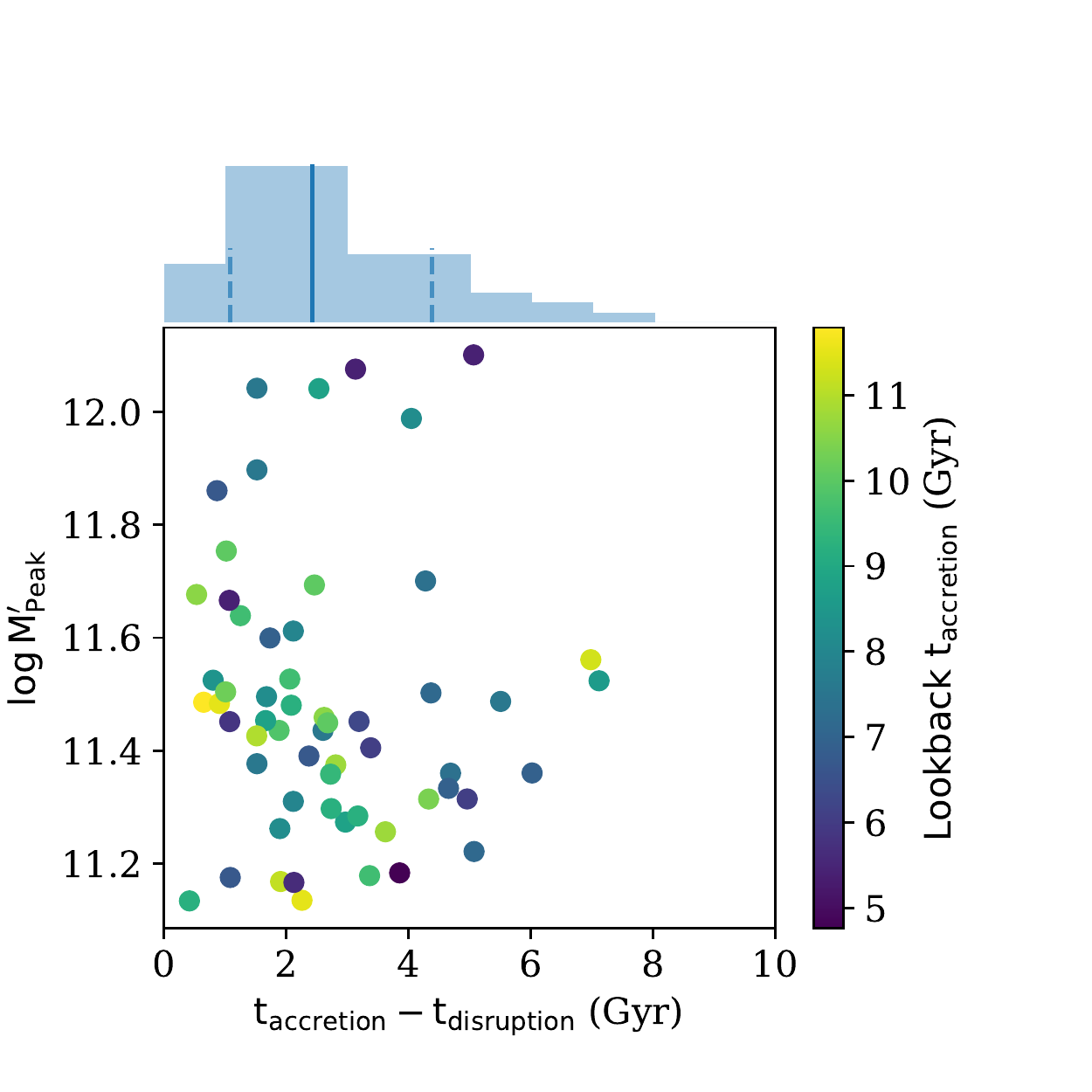}
    \caption{For the disrupted massive satellites, we estimate the merger time ($\mathrm{t_{accretion}-t_{disruption}}$) as a function of the peak mass ($M_{\rm peak}^{\prime}$) of the satellite. The histogram above describes the distribution of the merger time, while the points are coloured according to the their lookback time of accretion $\mathrm{t_{accretion}}$. The solid vertical line is the median, while the dashed vertical lines represent the 16th and 84th percentiles.}
    \label{app:fig1}
	\end{center}
\end{figure}

Evidence suggests that the time of disruption of M31's massive accreted progenitor was $\sim 2-3$ Gyr ago \citep[][]{DSouza2018b,Hammer2018}. Using the ELVIS simulations, we can constrain the approximate time of its accretion, i.e., its entry into M31's virial halo. In Fig. \ref{app:fig1}, we estimate the merger time of the massive accreted progenitors, i.e., the difference between the time of their accretion and the time of their disruption. We find that the median (mean) time of merger of the massive progenitors is $\sim$2.5 Gyr, with the lower 16 and higher 84 percentiles at 1 and 4.3 Gyr respectively. Furthermore, we find that massive progenitors which were accreted early fall in and merge rather quickly. These merger time scales are consistent with those derived by \cite{Kitzbichler2008}.  This suggests that M31's massive accreted progenitor was accreted between 5 and 6 Gyr ago.
 
\section{Selection of subhaloes hosting classical dwarfs}
\label{Appendix:2}
In this work, we have studied the infall of subhaloes along with a massive progenitor. However, we only observe dwarf satellite galaxies. In order to understand how our results might change with only satellites, we implement a simplified scheme to determine which subhaloes might host dwarf satellite galaxies ($\mathrm{M_{*}}> 10^{5} \mathrm{\msun}$).

Predicting the stellar mass content of low mass dark matter haloes is extremely challenging. However, in recent years, much progress have been made with models \citep[e.g.][]{Bullock2000,Somerville2002,Kravtsov2004,Koposov2009} now accounting for a variety of astrophysical processes including the suppression of star formation due to photoionisation, stellar feedback due to supernovae/galactic-winds as well as tidal stripping and disruption of subhaloes due to dynamical friction and the potential of the central baryonic disk. Although cosmological hydrodynamical simulations \citep[e.g.][]{Sawala2016,Simpson2018,Garrison2019} have recently attempted to reproduce these physical processes in a self-consistent way, the exact details of these processes are difficult to constrain with the limited data that are available, resulting in differences in the predictions of the slope and scatter in the stellar mass-halo mass relationship \citep{Digby2019}. Dwarf galaxy counts of nearby MW-mass galaxies argue for a significant scatter in the stellar mass-halo mass relationship \citep{Garrison2017a, Smercina2018}. A number of common elements are shared by all models incorporating astrophysical processes. First, in order to explain the `missing satellite' conundrum \citep{Klypin1999,Moore1999}, models suggest that classical dwarfs are those subhaloes which assembled a substantial fraction of their mass before reionization, and thus before the onset of photoionisation suppression. Furthermore, these massive subhaloes capable of hosting dwarf galaxies continue to form stars at a nearly constant rate prior to infall into a MW-mass halo, consistent with constraints from resolved star formation histories \citep{Weisz2014}.

\begin{figure}
	\begin{center}
	\includegraphics[width=0.8 \columnwidth]{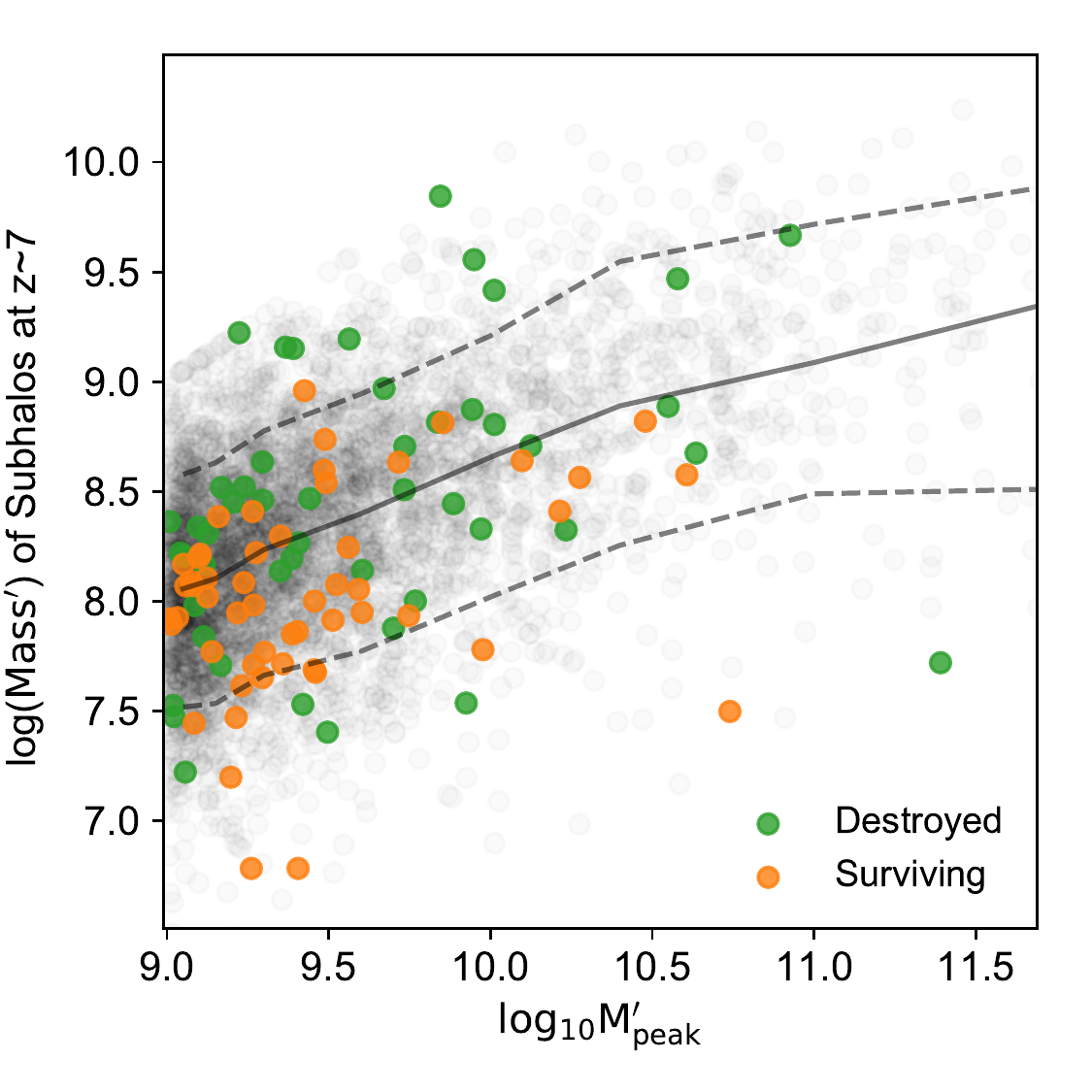}
    \includegraphics[width=0.8 \columnwidth]{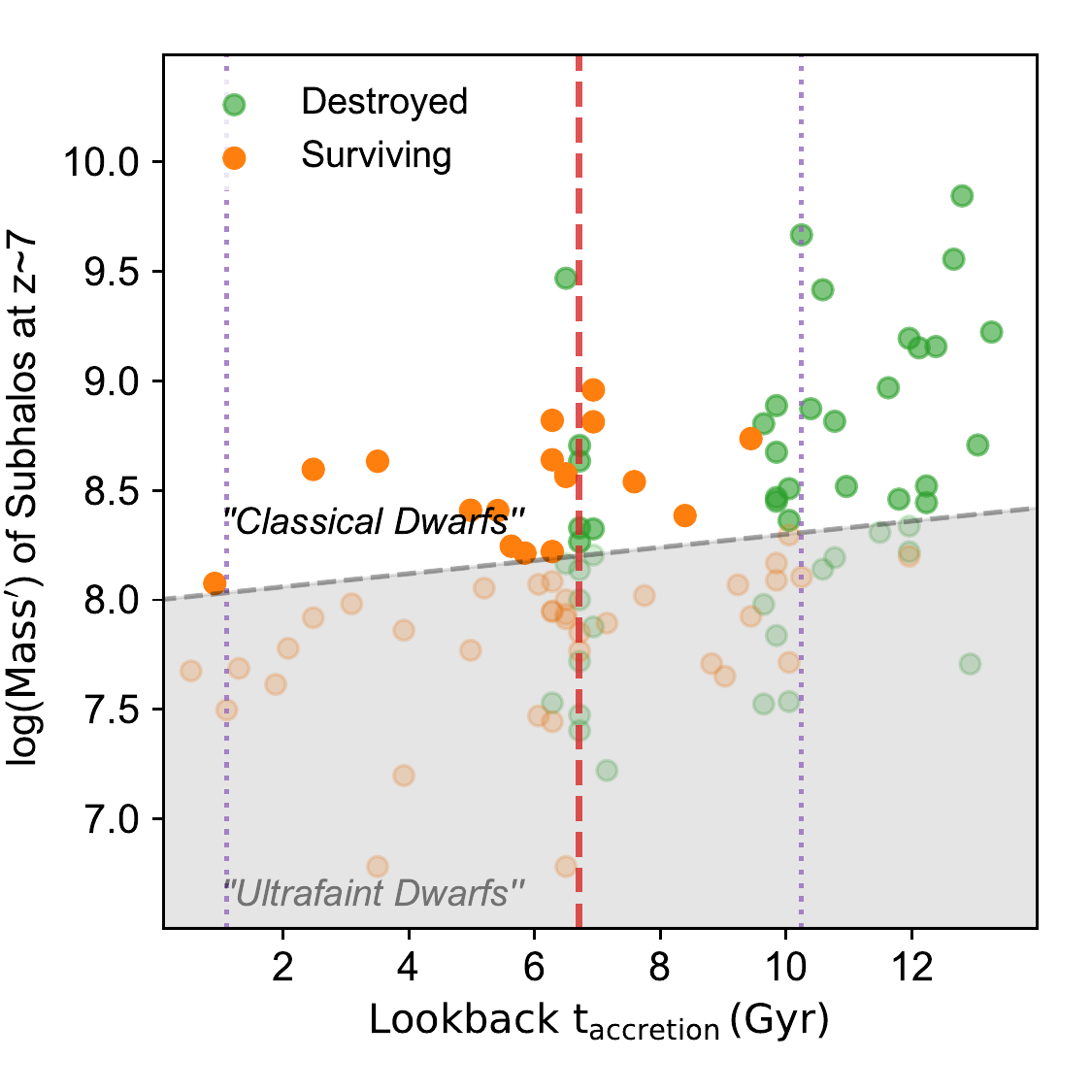}
    \caption{Top: The maximum mass of the subhaloes before $z\sim7$ as a function of $\mathrm{M^{\prime}_{peak}}$. The green and orange points indicate `destroyed' and `surviving' subhaloes of the MW-mass halo `iLincoln'. We also indicate the maximum mass at $z\sim7$ of all subhaloes of all the MW-mass haloes considered. The solid line indicates the running median as function of $\mathrm{M^{\prime}_{peak}}$, while the dashed lines indicate the 10th and 90th percentile. Bottom: The maximum mass of the subhaloes before $z\sim7$ as a function of their infall time for the MW-mass halo `iLincoln'. The dashed-line indicates our fiducial criterion for separating classical and ultra-faint dwarfs.}
    \label{app:fig2}
	\end{center}
\end{figure}

In order to develop an intuition of how these two principles might affect the proportion of classical dwarfs brought in by a massive accreted merger, we implement them in the following way. We assume that the present day stellar mass of the subhalo is a) strongly proportional to its virial mass at the end of reionization ($z\sim7$) and b) is a weak function of its infall time \citep[e.g.][]{Kravtsov2004}. The characteristic mass, the mass at which haloes on average lose half of their baryons due to photoionisation, at $z\sim7$ is $\sim \mathrm{M_{c}} \sim 10^{8} \mathrm{\msun}$ \citep{Okamoto2008}. Haloes more massive than this characteristic mass will continue to grow after reionization $z<7$ increasing its stellar mass by not more than 0.9 dex \citep{Weisz2014}, while we assume that less massive haloes will not be able to accrete gas. In the top panel of Fig. \ref{app:fig2}, we plot the maximum mass of the subhalo before $z\sim7$ as a function of $\mathrm{M_{peak}^{\prime}}$ for the MW-mass halo `iLincoln'. For a given $\mathrm{M^{\prime}_{peak}}$, there is a large scatter ($\sim$ 1 dex) in the maximum mass of the subhalo before $z\sim7$, which will lead to a similar or larger scatter in the stellar mass of the subhalo. In the bottom panel of Fig. \ref{app:fig2}, we plot the maximum mass of the subhaloes before $z\sim7$ as a function of their infall time. We adopt a fiducial criterion to separate classical from ultra-faint dwarfs, indicated by the dashed line in the bottom panel of Fig. \ref{app:fig2}; the slope of the line accounts for the fact that subhaloes above a characteristic mass can continue grow between reionization and infall into a MW-mass halo.

\begin{figure}
	\begin{center}
	\includegraphics[width=0.8 \columnwidth]{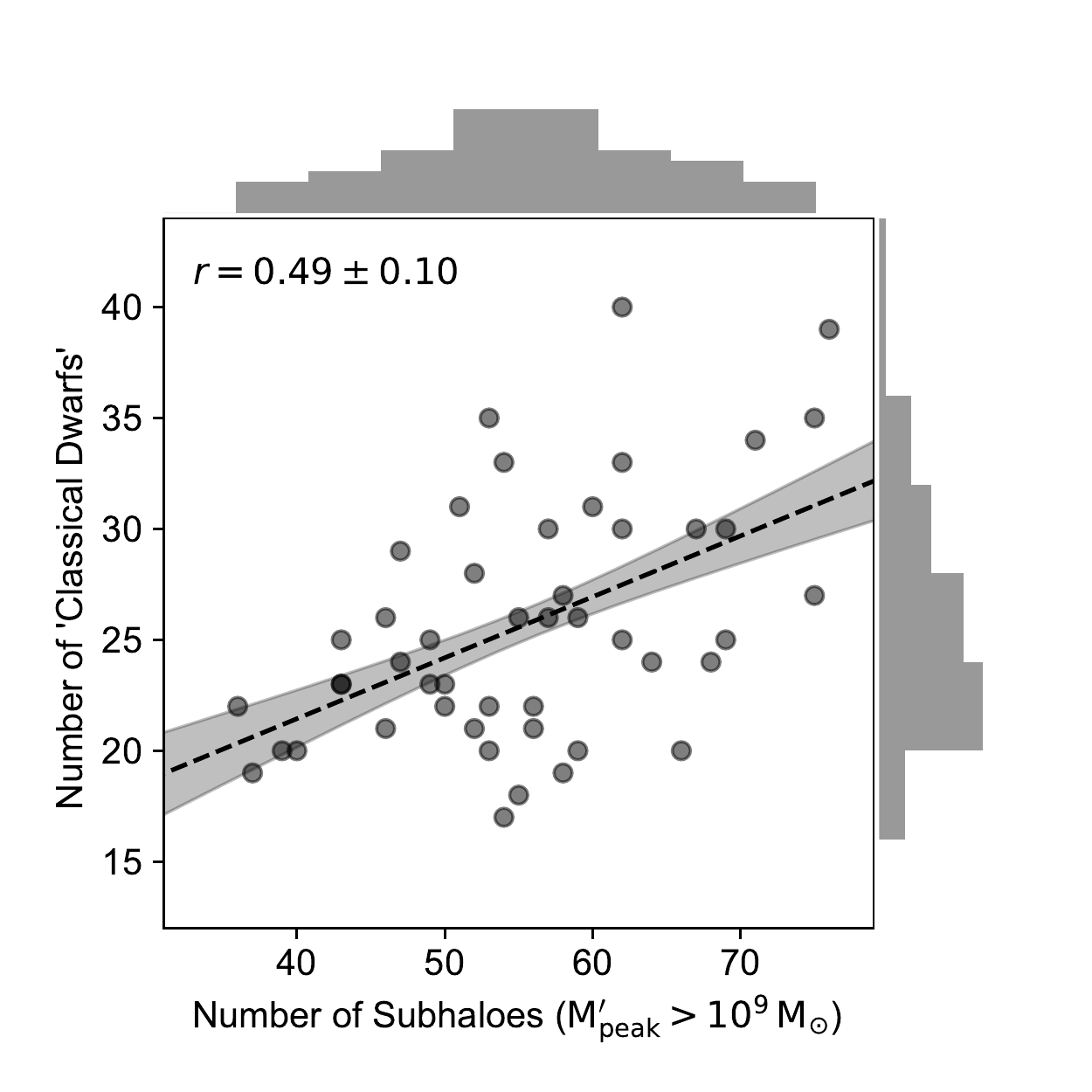}
    \caption{The number of surviving `Classical Dwarfs' vs the number of surviving subhaloes ($\mathrm{M^{\prime}_{peak}} > 10^{9} \msun$) found within the virial radius in our 48 MW-mass haloes of the ELVIS simulations by implementing the semi-analytical scheme outlined in Appendix \ref{Appendix:2}. The dashed line represents a straight line fit through the data, while the shaded region represent the 1-$\sigma$ uncertainty of the fit.}
    \label{app:fig3}
	\end{center}
\end{figure}

Using such a semi-analytical framework, we determine which of the infalling subhaloes of our 48 MW-mass haloes which survive upto $z=0$ are capable of hosting `classical dwarfs'. In Fig. \ref{app:fig3}, we plot the number of subhaloes capable of hosting classical dwarfs versus the number of subhaloes. We find that the number of subhaloes capable of hosting classical dwarfs scales with the total number of subhaloes with a $\sim$20\% scatter in the relation. The median number of subhaloes hosting classical satellites is $\sim$ 25. In Fig. \ref{fig:fig30}, we demonstrate that the clustering of the infall time of subhaloes hosting `classical satellites' and all subhaloes around massive accretions are the same.

\bsp	
\label{lastpage}
\end{document}